\documentclass[aps,prd,twocolumn,amsmath,amssymb,amsfonts,nofootinbib,superscriptaddress]{revtex4-1}
\usepackage[utf8]{inputenc}
\usepackage{graphicx}% Include figure files
\usepackage{subfigure}
\usepackage{datatool}
\usepackage[scientific-notation=true]{siunitx}
\usepackage[dvipsnames]{xcolor}
\usepackage{lineno}
\usepackage{xspace}
\usepackage{orcidlink}
\usepackage{url}

\usepackage{breakurl}
\usepackage{lineno}
\usepackage{longtable}
\usepackage[shortlabels]{enumitem}
\usepackage[capitalise]{cleveref}
\def\gwh{GW\xspace}
\def\cwh{CW\xspace}

\def\tcw{tCW\xspace}
\def\cws{CWs\xspace}
\def\gws{GWs\xspace}

\def\et{ET\xspace}
\def\nemo{NEMO\xspace}

\def\ce{Cosmic Explorer\xspace}
\def\dm{DM\xspace}
\def\dmh{DM\xspace}
\def\pbh{PBH\xspace}
\def\pbhs{PBHs\xspace}

\def\bbh{binary black hole\xspace}

\def\bns{binary neutron star\xspace}

\def\fh{\emph{frequency-Hough}\xspace}
\def\lvk{LIGO, Virgo and KAGRA\xspace}

\def\ifos{interferometers\xspace}
\def\hfgw{HFGW\xspace}

\def\mf{matched filtering\xspace}
\def\psd{power spectral density\xspace}
\def\nn{\nonumber}
\def\gc{Galactic Center\xspace}

\def\snr{signal-to-noise ratio\xspace}
\def\emri{EMRI\xspace}
\def\emris{EMRIs\xspace}

\newcommand{\bea}{\begin{eqnarray}}
\newcommand{\eea}{\end{eqnarray}}
\newcommand{\be}{\begin{equation}}
\newcommand{\ee}{\end{equation}}
\newcommand{\thetathr}{\theta_\text{thr}}

\newcommand{\avgVT}{\ensuremath{\left\langle VT \right\rangle}}
\newcommand{\avgVTtot}{\avgVT_{\rm tot}}

\newcommand{\TFFT}{T_\text{FFT}}
\newcommand{\fgw}{f_\text{gw}}

\newcommand{\fdotmax}{\dot{f}_{\rm max}}
\newcommand{\fdot}{\dot{f}}
\newcommand{\Tobs}{T_\text{obs}}
\newcommand{\fpbh}{f_\text{PBH}}

\newcommand{\ftilde}{\tilde{f}}
\newcommand{\ecc}{\epsilon}
\newcommand{\fsup}{f_\text{sup}}
\newcommand{\fmone}{f(m_1)}
\newcommand{\fmtwo}{f(m_2)}
\newcommand{\Mc}{\mathcal{M}}

\newcommand{\fmin}{f_\text{min}}
\newcommand{\fmax}{f_\text{max}}

\newcommand{\msun}{\ensuremath{M_\odot}\xspace}
\newcommand{\crthr}{CR_\text{thr}}

\newcommand{\mpbh}{M_\text{PBH}}
\newcommand{\dmax}{d_{\text{max}}}

\newcommand{\tcr}{\textcolor{red}}
\renewcommand{\tcr}[1]{#1}

\graphicspath{{./figures/}}

\begin{document}

\title{Prospects for detecting asteroid-mass primordial black holes in extreme mass-ratio inspirals with continuous gravitational waves}

% \title{Searches for continuous gravitational waves from extreme mass ratio inspirals \\ could detect asteroid-mass primordial black holes}

\author{Andrew L. Miller\,\orcidlink{0000-0002-4890-7627}}
\email{andrew.miller@nikhef.nl}
\affiliation{Nikhef -- National Institute for Subatomic Physics,
Science Park 105, 1098 XG Amsterdam, The Netherlands}
\affiliation{Institute for Gravitational and Subatomic Physics (GRASP),
Utrecht University, Princetonplein 1, 3584 CC Utrecht, The Netherlands}

\date{\today}

\begin{abstract}

Despite decades of research, the existence of asteroid-mass primordial black holes (PBHs) remains almost completely unconstrained and thus could still comprise the totality of dark matter (DM). In this paper, we show that standard searches for continuous gravitational waves -- long-lived, quasi-monochromatic signals -- could detect extreme mass-ratio inspirals of asteroid-mass PBHs in orbit around a stellar-mass companion using future gravitational-wave (GW) data from Einstein Telescope (ET) and the Neutron Star Extreme Matter Observatory (NEMO). We evaluate the robustness of our projected constraints against the eccentricity of the binary, the choice of the mass of the primary object, and the GW frequency range that we analyze. Furthermore, to determine whether there could be ways to detect asteroid-mass PBHs using current GW data, we quantify the impact of changes in current techniques on the sensitivity towards asteroid-mass PBHs. We show that methods that allow for signals with increased and more complicated frequency drifts over time could obtain much more stringent constraints now than those derived from standard techniques, though at slightly larger computational cost, potentially constraining the fraction of DM that certain asteroid-mass PBHs could compose to be less than one with current detectors.
\end{abstract}

\maketitle

\section{Introduction}

The primordial black hole (\pbh) hypothesis may explain a variety of puzzling cosmological observations, including the purported existence of \dm \cite{Hawking:1971ei,Chapline:1975ojl}, \lvk \bbh mergers ~\cite{aasi2015advanced,acernese2014advanced,Abbott:2016blz,TheLIGOScientific:2016pea,Abbott:2016nmj,Abbott:2017vtc,Abbott:2017oio,Abbott:2017gyy,LIGOScientific:2018mvr,LIGOScientific:2020stg,Abbott:2020uma,Abbott:2020khf,Abbott:2020tfl,Abbott:2020mjq}, excess microlensing events \cite{MACHO:1996qam}, the missing satellite problem \cite{DES:2015zwj}, and anisotropies in the cosmic microwave background \cite{Bullock:2017xww}, without introducing physics beyond the standard model \cite{Carr:2019kxo}. \pbhs could have formed from overdensities in the very early in the universe \cite{Khlopov:2008qy} with masses \tcr{given by} \cite{Villanueva-Domingo:2021spv}:

\begin{equation}
    \mpbh\sim 5\times 10^4\msun \left(\frac{t}{1\text{ s} }\right),
\end{equation}
where $t$ is the time from the Big Bang that \pbhs formed.
% Depending on when they formed in the early universe, primordial black holes (\pbhs) could take on virtually any mass ranging from $\sim 10^{-22}\msun$ to $10^9\msun$ \cite{Green:2020jor}. 
Such a wide mass range motivates the need to consider a variety of \pbh formation mechanisms and experiments to detect the presence of such objects. 

% using \gwh upper limits on merging rates of solar-mass black holes \cite{}, Hawking evaporation of ultralight \pbhs \cite{}, microlensing of star light by planetary-mass and solar-mass \pbhs, accretion of interstellar gas onto \pbhs \cite{}, and dynamical considerations of potentailly increased star velocities in the presence of \pbhs \cite{}, a par

If \pbhs exist, they could compose a fraction or the totality of dark matter (\dm) and particular observation signatures \cite{Belotsky:2014kca,Clesse:2017bsw,Belotsky:2018wph}. While the fraction of \dm that \pbhs could compose, $\fpbh$, has been heavily constrained across a wide parameter space, one particular regime, asteroid-mass ( $10^{-16}\msun\lesssim \mpbh\lesssim 10^{-12}\msun$) \pbhs, has been able to evade almost all constraints. In this regime, \pbhs would be too heavy to avoid evaporating\footnote{AMEGO, however, may be able to detect evaporating asteroid-mass \pbhs emitting MeV photons \cite{Ray:2021mxu}. }, but too light to amplify the flux from stars that they lens to a high-enough level detectable by microlensing experiments. In particular, if \pbhs have masses of $\lesssim 10^{-10}\msun$, their Schwarzschild radii are approximately equal to, or less than, the wavelength of the light that is being lensed \cite{Green:2020jor}, resulting in a strong suppression of the amplification induced by the lens \cite{Ulmer:1994ij}\footnote{In principle, the Hyper Suprime-Cam (HSC) experiment \cite{Aihara:2017paw,Aihara:2017tri} can probe $\fpbh\lesssim 1$ for lensing masses down to $10^{-12}\msun$ \cite{Niikura:2017zjd}, although if the light source has a larger size than assumed in \cite{Smyth:2019whb}, the constraints are almost completely relaxed between $\sim[10^{-12},10^{-11}]\msun$ \cite{Smyth:2019whb}.}.

The inability to obtain constraints in this mass regime motivates the need to consider alternative ways of probing asteroid-mass \pbhs.

One promising avenue could be to search for femtolensing of seconds-long gamma ray bursts (GRBs) \cite{gould1992femtolensing}. In femtolensing, light rays from a GRB would traverse different path-lengths around the \pbh lens and thus result in an interference pattern
% , i.e. \tcb{oscillations in the GRB energy spectra from photons, since different photons with different energies will experience the same time delay} 
\cite{Davidson:2016uok}.
In particular, a lack of observed femtolensing events of GRBs observed by \emph{Fermi} has led to constraints on $\fpbh\lesssim 0.1$ between $\sim [10^{-16},10^{-14}]\msun$ \cite{Barnacka:2012bm}, although the assumption of point-mass GRB sources has been questioned: if the GRB source has a finite size, photons would be emitted from different locations on the source, which would all follow different paths as they bend around the lens, potentially masking any oscillations visible in the energy spectrum \cite{Davidson:2016uok}.  Additionally, the large sizes of GRBs relative to the lensing mass in the lensing plane make most unsuitable for such searches, relaxing most projected constraints \cite{Katz:2018zrn}. Thus, observations of small-size GRBs are needed to produce realistic constraints in the asteroid-mass regime.

Another avenue could be to look at microlensing of light from x-ray pulsars by asteroid-mass \pbh lenses, because (1) wave-optics effects are negligible for $\mpbh\gtrsim 10^{-13}\msun$ \cite{gould1992femtolensing}, and (2) pulsars are compact objects, which reduces the impact of finite-source size on the lensing of light compared to using extended sources of light \cite{Hickox:2004fy}. While it is not currently possible to constrain \pbhs using RXTE \cite{Levine:1996du}, future observations with eXTP \cite{eXTP:2018anb} may constrain $\fpbh\lesssim 1 $ between $\sim$ [few $\times 10^{-16}$, few $\times 10^{-13}]\msun$ \cite{Bai:2018bej}, reaching at best $\fpbh\simeq0.1$ at $2\times 10^{-15}\msun$ (see Fig. 5 of \cite{Bai:2018bej}).

Even if eXTP could obtain such constraints in the future, a key mass regime still remains unconstrained that spans almost an order of magnitude, $\sim[10^{-13},10^{-12}]\msun$, which lies in between the eXTP and Subaru/HSC constraints. 

One possible probe of this mass regime lies in attempting to detect \gws from very slowly inspiraling extreme mass-ratio binaries, consisting of an ordinary compact object with a mass $\mathcal{O}(\msun)$ and a significantly lighter asteroid-mass compact object, i.e. a \pbh. Such so-called extreme-mass ratio inspirals (\emris) are usually considered in the context of future spaced-based \gwh detectors, in which an ordinary compact object of $\mathcal{O}(\msun)$ inspirals around and then plunges into a supermassive black hole \cite{Babak:2017tow}, though others have considered so-called ``mini-\emri'' systems, composed of $\mathcal{O}(1-10)\msun$ ordinary compact object with a planetary-mass exotic compact object orbiting around it, that would be visible in current ground-based detectors \cite{Guo:2022sdd}. 

Additionally, asteroid-mass \pbhs could be constrained via their imprints on a stochastic \gwh background. If \pbhs are generated from overdensities in the early universe, they will lead to scalar-induced \gws, which, if not detected by future space-based detectors, such as LISA, would also stringently constrain asteroid-mass \pbhs \cite{Yuan:2019udt,Meng:2022low}.

Here, consider solely the inspiral portion of \emri systems, specifically at frequencies at which the \gwh signal appears to be almost monochromatic and ever-lasting. Thus, the signal falls into the category of ``continuous waves'' (\cws), for which extensive method development has taken place over the last few decades within and outside of the \lvk collaborations \cite{Sieniawska:2019hmd,Tenorio:2021wmz,Riles:2022wwz,Piccinni:2022vsd,Miller:2023qyw,Wette:2023dom}. However, by only analyzing the inspiral portion of the lifetime of the \emri system, we pay a price in sensitivity: while searches for merging black holes reach could reach out to at least Gpc, we could, at best, detect systems at the scale of the Galaxy, $\mathcal{O}$(kpc) \cite{Miller:2021knj}. Despite the small distance reach of \gwh searches to such systems, we are still able to project stringent constraints in the asteroid-mass \pbh regime.

In this paper, we make a number of considerations regarding the prospects for detection of asteroid-mass \pbhs using \cwh search techniques. While previous works have made specific choices of $m_1,m_2$ and ignored eccentricity in the context of current \cwh searches \cite{Miller:2021knj,KAGRA:2022dwb}, and have not yet been able constrain $\fpbh\leq 1$, we now generalize our thinking to see how constraints on the fraction of \dm that \pbhs could compose change as a function of (1) \gwh detector choice, (2) \cwh search parameters ($\TFFT,\fdotmax$, etc.), (3) source parameters ($m_1,m_2$, eccentricity), and (4) \gwh frequency bands that are analyzed. Our considerations indicate that searches for \cws can make significant contributions to constraining the asteroid-mass \pbh regime, especially if some analysis methods are tuned to consider \pbh \emris, \emph{and} even if no changes are made to existing analysis methods.

\section{Signal model}

Two compact objects in orbit around their center of mass will emit \gws as they approach each other. Equating the orbital energy loss with \gwh power, we can obtain the rate of change of the frequency over time, i.e. the spin-up $\fdot$, in the quasi-Newtonian limit (i.e. far from merger) \cite{maggiore2008gravitational}:

\begin{align}
    \dot{f}_{\rm gw}&=\frac{96}{5}\pi^{8/3}\left(\frac{G\Mc}{c^3}\right)^{5/3} \fgw^{11/3}\equiv k \fgw^{11/3} \nonumber \\
    &\simeq 9.83\times 10^{-11}\text{ Hz/s} \left(\frac{\Mc}{10^{-6}M_\odot}\right)^{5/3}\left(\frac{\fgw}{50\text{ Hz}}\right)^{11/3},
    \label{eqn:fdot_chirp}
\end{align}
where $\Mc\equiv\frac{(m_1m_2)^{3/5}}{(m_1+m_2)^{1/5}}$ is the chirp mass of the system, $\fgw$ is the \gwh frequency, $c$ is the speed of light, and $G$ is Newton's gravitational constant.  \tcr{Throughout the paper, unless we explicitly say we are considering eccentricity, the results assume quasi-circular orbits.}

To obtain the signal frequency evolution $\fgw(t)$ over time, we integrate \cref{eqn:fdot_chirp} with respect to time $t$:
\begin{equation}
\fgw(t)=f_0\left[1-\frac{8}{3}kf_0^{8/3}(t-t_0)\right]^{-\frac{3}{8}}~,
\label{eqn:powlaws}
\end{equation}
where $t_0$ is a reference time for the \gwh frequency $f_0$. 

In most of this paper, we consider systems far from merger with sufficiently low $\Mc\lesssim 10^{-5}\msun$, which means we can binomially expand \cref{eqn:powlaws}, which corresponds to $\dot{f}_{\rm gw}(t-t_0)\ll f_0$:

\begin{equation}
\fgw= f_0+\fdot_{\rm gw}(t-t_0).\label{eqn:ftay}
\end{equation}
Our model for \gws from inspiraling systems is thus a sinusoid whose frequency slowly varies over time.

The amplitude $h_0(t)$ of the \gwh signal also evolves with time \cite{maggiore2008gravitational}:

\begin{align}
h_0(t)&=\frac{4}{d}\left(\frac{G \Mc}{c^2}\right)^{5/3}\left(\frac{\pi \fgw(t)}{c}\right)^{2/3} \nonumber \\
&\simeq 1.61\times 10^{-25}\left(\frac{1\text{ pc}}{d}\right)\left(\frac{\Mc}{10^{-6}\msun}\right)^{5/3}\left(\frac{\fgw}{50\text{ Hz}}\right)^{2/3},
\label{eqn:h0}
\end{align}
where $d$ is the luminosity distance to the source.

Inverting \cref{eqn:powlaws}, we can also write down an expression for the time the signal spends between two frequencies:

\begin{equation}
    \Delta t=-\frac{3}{8}\frac{\fgw^{-8/3}-f_0^{-8/3 }}{k},\label{eqn:dt-bt-2-freqs}
\end{equation}
which, in the limit that $\fgw\rightarrow f_{\rm isco}$, where $f_{\rm isco}$ is the frequency at the innermost stable circular orbit (ISCO), and $f_{\rm isco}\gg f_0$, determines the time to merger $t_{\rm merg}$

\begin{align}
    t_{\rm merg}&\simeq\frac{5}{256}\left(\frac{1}{\pi f_0}\right)^{8/3}\left(\frac{c^3}{G\Mc}\right)^{5/3}\nonumber \\
    &\simeq 6000\text{ years} \left(\frac{50\text{ Hz}}{f_0}\right)^{8/3} \left(\frac{10^{-6}\msun}{\Mc}\right)^{5/3}.
    \label{eqn:tmerg}
\end{align}

\section{Searches for continuous waves}\label{sec:cwsearch}

\subsection{Background}
Multiple groups search for \cws emitted from anywhere in the sky \cite{KAGRA:2022dwb,Steltner:2023cfk}. The parameter space \tcr{for all-sky searches for isolated neutron stars typically} has four dimensions: $f,\fdot,\alpha,\delta$, where $\alpha,\delta$ refer to the sky position of the source. Despite such a simplistic signal model, these searches cannot be performed fully coherently, since each sky position has to be targeted individually, and the computational cost of such analyses scales with $\Tobs^6$ \cite{Prix:2009oha}. This means that \emph{semi-coherent} \cwh search techniques must be employed, which divide the data into smaller chunks of length $\TFFT\ll \Tobs$, in order to both save computational costs ($\propto \Tobs^2\TFFT^4$ now) and to increase robustness against theoretical uncertainties in the \gwh search (e.g. phase coherence could be lost over such long observation times). Computational limitations also require that a fixed range of $\fdot$ be searched over, since the number of steps in the $\fdot$ grid scales with \tcr{$\TFFT\Tobs$}\cite{Prix:2009oha}. 

\begin{figure}
    \centering
    \includegraphics[width=0.49\textwidth]{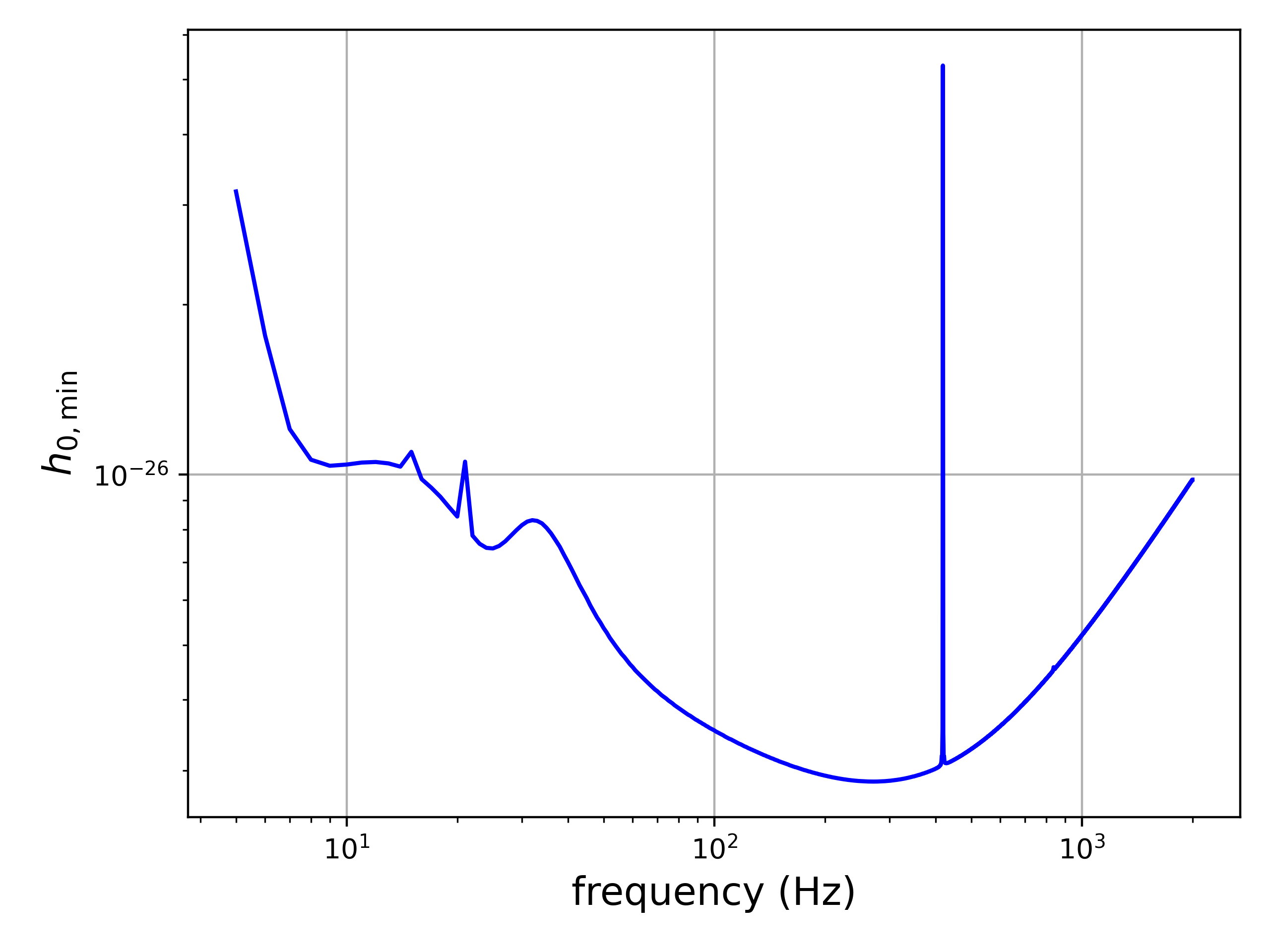}
    \caption{Synthetic upper limits obtained using \cref{eqn:h0min} in each 1 Hz band using an \et sensitivity curve and the following analysis parameters: $\TFFT=1.5$ days; $\Tobs=1$ year; $\Gamma=0.95$, $\crthr=5$, and $\thetathr=2.5$ .}
    \label{fig:h0min-synth}
\end{figure}

When searches for \cws do not find a significant candidate, upper limits on the minimum detectable \gwh amplitude are produced, usually averaged over the sky but not always \cite{KAGRA:2022dwb,Dergachev:2024knd}, as a function of frequency. These limits are derived either through injecting fake signals and recovering them with a particular method, or through analytic / data-driven procedures that encapsulate the properties of the noise while producing conservative constraints with respect to those that could have been obtained through injections \cite{Miller:2020vsl,LIGOScientific:2021odm, KAGRA:2022dwb,Miller:2024fpo,Miller:2024jpo}. An example set of sky-averaged synthetic upper limits that we would obtain on the \gwh amplitude $h_{0,\rm min}$ using the \et \psd is shown \cref{fig:h0min-synth}, which will be the starting point for our calculations of projected constraints on \pbhs.

\subsection{Limitations of current searches and constraints on asteroid-mass \pbhs}

In practice, when performing an all-sky search, preference is usually given to negative values of $\fdot$, since the proposed targets of such searches, lumpy isolated neutron stars, should be spinning down, and thus $|\fdot|\lesssim 10^{-9}$ Hz/s. The maximum positive $\fdot$ searched for limits the extent to which we can use \cref{eqn:ftay} to approximate \cref{eqn:powlaws}. Thus, efforts to constrain \pbh abundance cannot typically leverage upper limits at frequencies above $\sim 100$ Hz (see Fig. 1 of \cite{Miller:2021knj}), unless $\Mc$ is sufficiently small, which, unfortunately, implies smaller $h_0$ (\cref{eqn:h0}). Furthermore, even though in principle \pbh abundance for \emri systems using results from \cwh searches can be constrained, Refs. \cite{Miller:2021knj,KAGRA:2022dwb} do not yet account for eccentricity of such binaries, and how that would impact the constraints, which we will do in \cref{sec:anacons}.

\section{Methodology}\label{sec:meth}

% \subsection{All-sky searches}

% \subsection{Directed searches}
% \subsection{How to set constraints on \pbh abundance}
Following \cite{Miller:2021knj,Miller:2024rca,Bhattacharya:2024pmp}, we compute the sensitivity to inspiraling \pbh binaries obtainable with semi-coherent techniques. In our case, we use a formula to compute the sensitivity of the \fh (a pattern recognition technique used in all-sky searches \cite{Astone:2014esa}) towards quasi-monochromatic sources that are sinusoidal within each $\TFFT$:

\begin{align}
h_\text{0,min}&\simeq
\Lambda\sqrt{\frac{S_n(\fgw)} {\TFFT^{1/2}\Tobs^{1/2}}} \nn \\
&\simeq 5.37\times 10^{-27} \left(\frac{1.5\text{ days}}{\TFFT}\right)^{1/4} \left(\frac{1\text{ year}}{\Tobs}\right)^{1/4} \nn \\ &\times
\left(\frac{S_n(\fgw=50\text{ Hz})}{3.58\times 10^{-49} \text{ Hz}^{-1}}\right)^{1/2} \left(\frac{\Lambda}{12.81}\right)
\label{eqn:h0min}
\end{align}
where $\Lambda=12.81$ \tcr{based on the official sensitivity of the \fh, see Eq. 67 of \cite{Astone:2014esa}}. \tcr{Note that $\TFFT=1.5$ days is an assumption: we do not know what computational resources we will have in the next-generation detector era, but as in \cite{Branchesi:2023mws}, we assume that we will be able to take longer ones than what we use now (e.g. $\TFFT=8192$ s at frequencies below 128 Hz \cite{Astone:2014esa}). Using $\TFFT=8192$ s would result in a factor of 2 worse sensitivity in $h_\text{0,min}$, thus a factor of 8 smaller on the rate density projected constraints presented below.}

Combining \cref{eqn:h0min} and \cref{eqn:h0}, we can obtain an expression for the luminosity distance reach to \pbh inspirals:

\begin{align}
    d(\fgw) &=\frac{1}{\Lambda} \sqrt{\frac{\TFFT^{1/2} \Tobs^{1/2}}{{S_n(f)}}}\left(\frac{G \mathcal{M}}{c^2}\right)^{5/3}\left(\frac{\pi \fgw}{c}\right)^{2/3} \nn \\
    &\simeq 30\text{ pc} \left(\frac{1.5\text{ days}}{\TFFT}\right)^{-1/4} \left(\frac{1\text{ year}}{\Tobs}\right)^{-1/4} \nn \\ &\times
\left(\frac{S_n(\fgw=50\text{ Hz})}{3.58\times 10^{-49} \text{ Hz}^{-1}}\right)^{-1/2} \left(\frac{\Lambda}{12.81}\right)^{-1} \nn \\ &\times \left(\frac{\Mc}{10^{-6}\msun}\right)^{5/3}\left(\frac{\fgw}{50\text{ Hz}}\right)^{2/3}
\label{eqn:dmax95}
\end{align}
\tcr{It is also worth quoting these quantities in method-independent parameters, particularly the \emph{sensitivity depth} $\mathcal{D}=\sqrt{S_n}/h_{0,\rm min}$ that would allow interested readers to, with their own methods, assess their sensitivity to asteroid-mass \pbhs. In terms of $\mathcal{D}$: }

\begin{align}
    d(\fgw) =& 30 \, \text{pc} \, \left(\frac{\Mc}{10^{-6} \, M_\odot}\right)^{5/3} \left(\frac{\fgw}{50 \, \text{Hz}}\right)^{2/3} \nonumber  \\
    \times& \left(\frac{\mathcal{D}}{111 \, \text{Hz}^{-1/2}}\right) \left(\frac{6 \times 10^{-25} \text{ Hz}^{-1/2} }{\sqrt{S_n}}\right)
\end{align}
After computing the luminosity distance reach, we can write down the space-time volume $\avgVT$:

\begin{equation}
    \avgVT = \frac{4}{3}\pi d(\fgw)^3 T,\label{eqn:vt}
\end{equation}
where the brackets indicate the expectation value of the space-time volume.
One part of this equation is the Euclidean volume of a sphere, since we do not need to consider any cosmological effects for nearby sources. The other piece, $T$, is: $T=$max$(\Tobs,\Delta T)$. \(\Delta T\) is the time spent by the binary system in a given frequency range $[f,f+\delta f]$: 
\begin{align}
  \Delta T &= \frac{5}{256}\pi^{-8/3}\left(\frac{c^3}{G\mathcal{M}}\right)^{5/3} \left[\fgw^{-8/3}-(\fgw+\delta f)^{-8/3 }\right] \nn \\ ~ %\\
  &\approx  310\text{ years} \left(\frac{10^{-6}\msun}{\mathcal{M}}\right)^{5/3} \left(\frac{\delta f}{1\text{ Hz}}\right) \left(\frac{\fgw}{50\text{ Hz}}\right)^{-11/3}
  % &\approx \frac{5}{96}\pi^{-8/3}\left(\frac{c^3}{G\mathcal{M}}\right)^{5/3} \delta f \fgw^{-11/3}
\label{eqn:deltaT}
\end{align}
where $\delta f$ is the spacing in frequency at which we evaluate $d(f)$. Note that we consider this formulation of $T$ because the source durations tend to greatly exceed the observation time, so we are sensitive, in one particular frequency bin, to multiple sources emitting \gws before and/or during $\Tobs$ \cite{Miller:2021knj,Bhattacharya:2024pmp}. \tcr{$\delta f$ is essentially the spacing in frequency at which the upper limits is computed in a real search, when such upper limits are computed with injections. It is not an infinitesimally small quantity, and  determines how ``monochromatic'' the upper limit at each frequency is. Physically, $\delta f$ contains overlapping \pbh inspirals that are all at the ``same'' frequency (within $\delta f$). As we show in \cref{app:deltaf}, it does matter how big or small this quantity is, because in the regime in which the source durations greatly exceed the observation time, we are sensitive, in $\delta f$, to multiple sources emitting \gws.}

Now, the number of binaries detectable at a given frequency is:

\begin{equation}
    N_{\rm bin}=\avgVT \mathcal{R},\label{eqn:nbin}
\end{equation}
where $\mathcal{R}$ is the formation rate density of binary \pbhs. Summing over all possible binaries detected at each frequency

\begin{equation}
N_{\rm bin}^{\rm tot} = \sum_{ i} N_{\rm bin} (f_i)~.
\label{eqn:ntot}
\end{equation}
and solving for $\mathcal{R}$, assuming no detection ($N_{\rm bin}^{\rm tot}<1)$, we arrive at

\begin{equation}
\mathcal{R}=\left(\sum_i \avgVT(f_i)\right)^{-1}.
\label{eqn:ratedenssolved}
\end{equation}
% These rate densities are a function of each chirp mass to which we could be sensitive.
We can equate the rate densities in 
\cref{eqn:ratedenssolved} to analytic models \cite{Clesse:2015wea,Clesse:2016vqa} for formation rate densities of \pbhs in the case of asymmetric-mass ratio binaries, with $m_2\ll m_1$:

 \begin{align}
\mathcal{R} &= 5.28 \times 10^{-7}\, \mathrm{kpc}^{-3} \mathrm{yr}^{-1} f_{\rm sup} f(m_1) f(m_2) \nn \\ &\times \left(\frac{m_1}{M_\odot}\right)^{-32/37} \left(\frac{m_2}{m_1}\right)^{-34/37} \left(f_{\rm PBH}\right)^{53/37}~,
\label{eqn:rate_asymm}
\end{align}
where $\fsup$ is suppression factor induced by the presence of nearby \pbhs that could break up binaries, $\fmone$ and $\fmtwo$ are the mass functions of the primary and secondary components of the binary, respectively, \tcr{and $\fpbh$ is the fraction of \dm that \pbhs could compose.}

To remain agnostic against particular \pbh mass functions and possible rate suppression factors $\fsup$, we quote all constraints in terms of an effective fraction $\ftilde$:

\begin{equation}
\ftilde \equiv \fpbh \left[\fsup\fmone\fmtwo\right]^{37/53},
\label{eqn:ftilde}
\end{equation}
% In contrast to \cite{Miller:2021knj,KAGRA:2022dwb}, we relax some of the strict criteria necessary to set upper limits on $\ftilde$ and use \cref{eqn:h0min} to generate synthetic upper limits $h_\text{0,min}(f)$ for different proposed detectors to forecast constraints on \pbh abundance.

It is important to note that \cwh searches will be sensitive only to $\Mc$, which is a particular combination of $m_1$ and $m_2$. Thus, when we interpret synthetic upper limits in terms of constraints on $\ftilde$, we are free to pick $m_2$ to be in the asteroid-mass range as long as $m_1$ is sufficiently heavy to obtain the same $\Mc$. Of course, such \emri systems may have high eccentricities, which we consider as well in the following sections.

\section{Projected constraints}\label{subsec:etsens}

% We forecast constraints on $\ftilde$ in Einstein Telescope (\et) and particular \hfgw detectors: the Neutron Star Extreme Matter Observatory (\nemo) \cite{} and Levitated Sensor Detector (\lsd) \cite{}

% \subsection{\et Sensitivity to extreme mass ratio inspirals (varying $m_1$ and $m_2$)}\label{subsec:etsens}

\et will provide unprecedented low-frequency sensitivity to \gws, enabling us to see much longer signals arising from the inspirals of two compact objects than currently possible with \lvk. Furthermore, the sensitivity across all frequencies will increase by approximately an order of magnitude, enabling numerous detections of \bbh and \bns mergers \cite{Maggiore:2019uih,Branchesi:2023mws}.

\begin{figure*}[ht!]
     \begin{center}
        \subfigure[ ]{%
            \label{fig:distreachet}

        \includegraphics[width=0.5\textwidth]{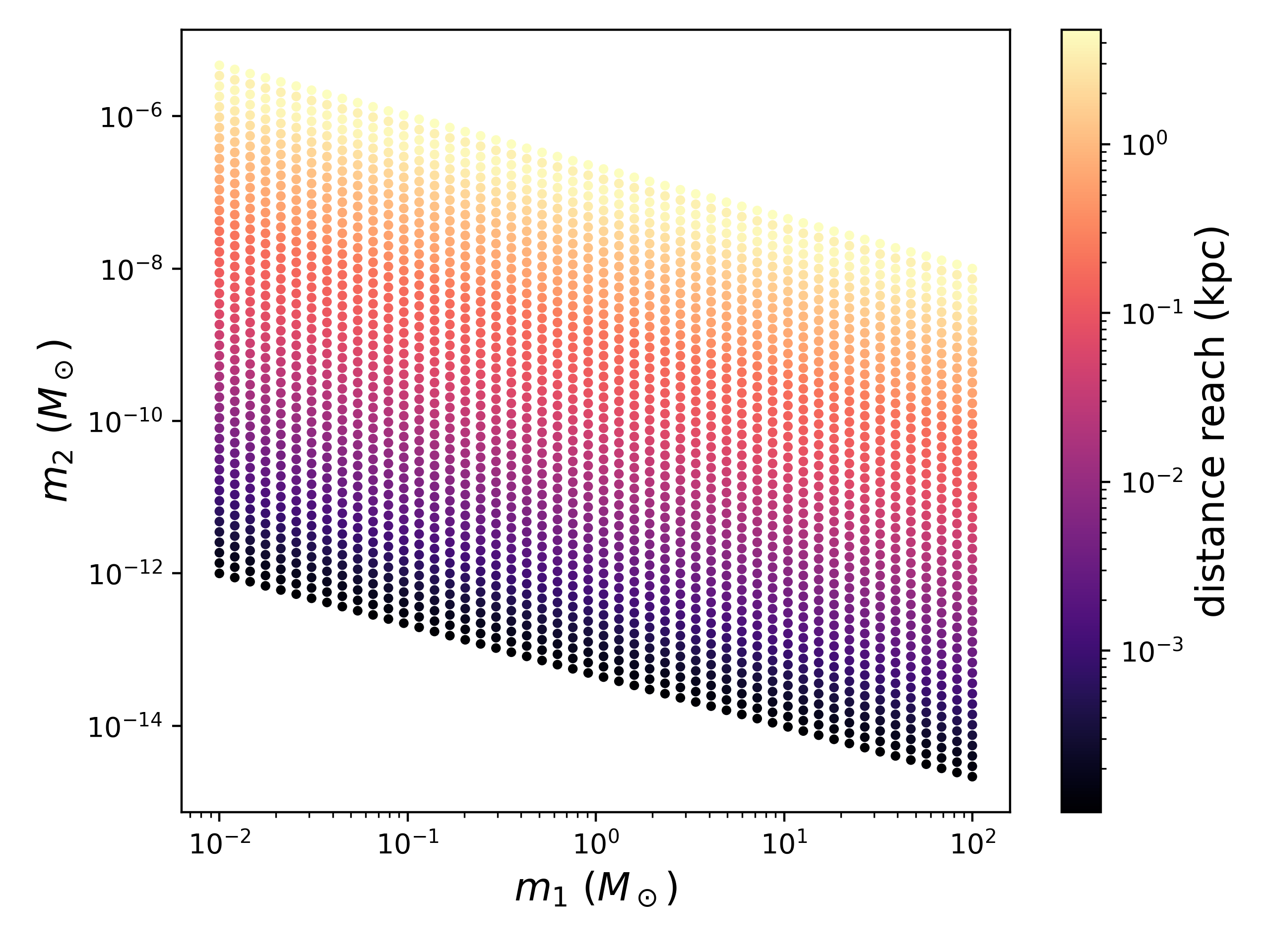}
        }%
        \subfigure[]{%
           \label{fig:fpbh-constr-et}
           \includegraphics[width=0.5\textwidth]{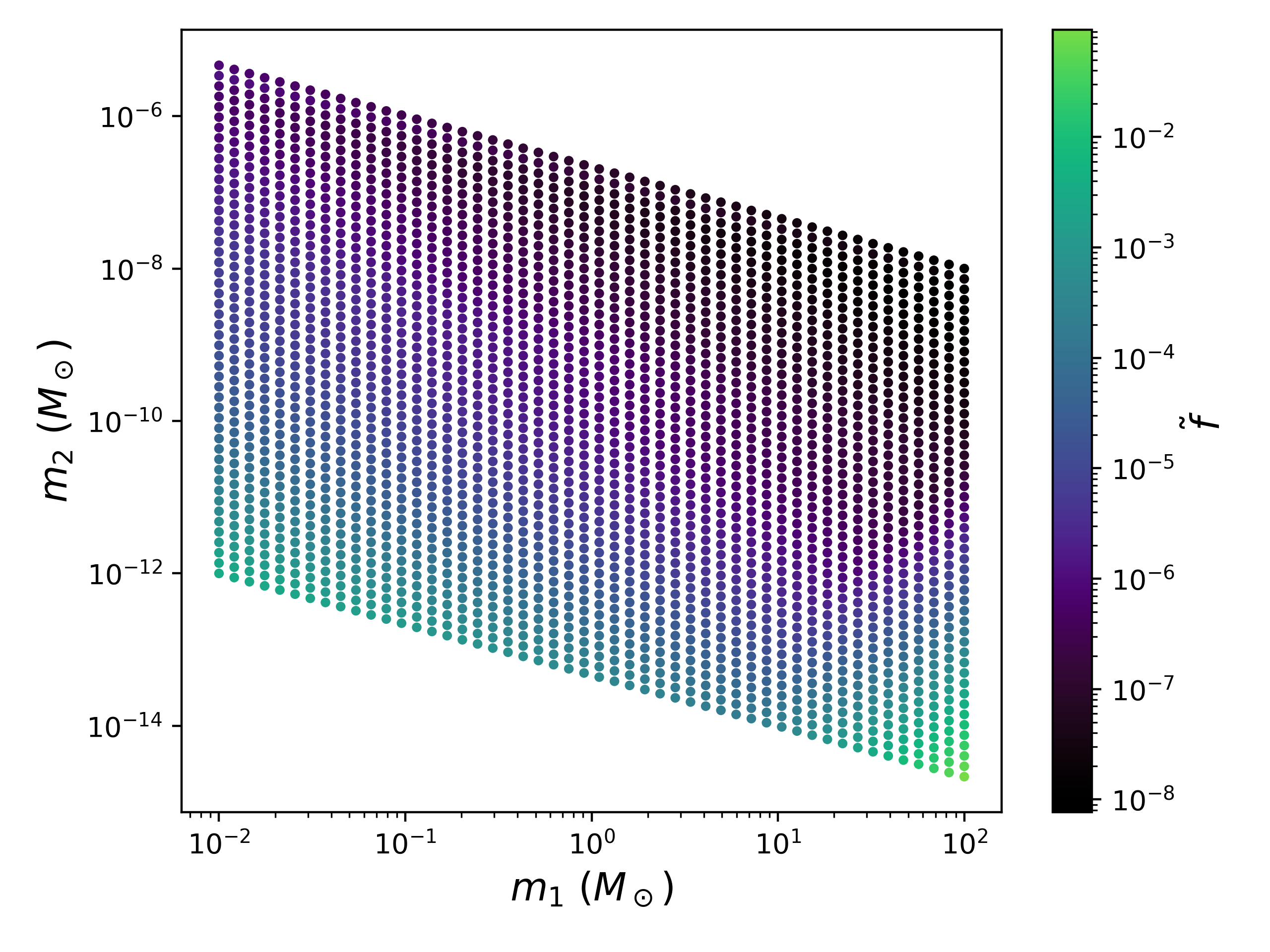}
        }\\ %  ------- End of the first row ----------------------%
    \end{center}
     \caption[]{Using \cref{eqn:dmax95} and the \et \psd curve, we have computed the expected luminosity distance reach (left) and $\ftilde$ (right) as a function of $m_1$ and $m_2$, enforcing the criteria that $\fdot\leq\fdotmax=10^{-9}$ Hz/s and that the linear approximation in \cref{eqn:ftay} holds. We have only plotted points in which $\ftilde<1$, and have assumed that the eccentricity is negligible, an assumption that will be relaxed later in \cref{sec:anacons}. }%
     \label{fig:dist-and-fpbh-m1-m2}
\end{figure*}

It is thus worth asking to what extent \et will be able to detect asteroid-mass \pbhs that could form in a binary system. Following the methodology outlined in \cref{sec:meth}, and considering a range of possible \emri systems, we compute the expected luminosity distance reach and constraints on $\ftilde$, as shown in \cref{fig:distreachet}. We have selected the maximum distance reach possible for each system to plot here, that is, selecting the particular frequency at which the distance reach is maximum. Additionally, we have assumed $\fdotmax<10^{-9}$ Hz/s, which was used in \cite{KAGRA:2022dwb}, and a frequency evolution given by \cref{eqn:ftay}. Though we make these plots in terms of $m_1$ and $m_2$, the distance reach depends primarily on $\Mc$, which is constant along the diagonal lines in these plots. When computing $\ftilde$, however, we use the contributions for \emri systems inspiraling at \emph{all} frequencies, not just the maximum one, as implied by \cref{eqn:ratedenssolved}. We evaluate the impact of the range of frequencies chosen on the $\ftilde$ constraint in \cref{sec:anacons}.

\section{Analysis considerations
}\label{sec:anacons}
Here, we consider what changes current and future \cwh analyses could make to enhance their sensitivity to asteroid-mass \pbhs.
% \hfgw detectors do not seem to be the most promising to probe asteroid-mass \pbhs. We therefore focus on ground-based \gwh detectors.

\subsection{Varying $\dot{f}_{\rm max}$}\label{subsec:varyfdot}

A major limitation of \cwh searches to probe asteroid-mass \pbhs is the linear signal model (\cref{eqn:deltaT}) and the $\fdotmax$ considered. These two criteria are linked: systems with chirp masses above a critical value will require us to use second- or third-order terms in \cref{eqn:ftay} to model them correctly. Furthermore, even among systems that do follow \cref{eqn:ftay}, $\fdotmax$ prevents higher frequencies from contributing to the sum in \cref{eqn:ratedenssolved}, which, as we will argue, degrades the constraint on $\ftilde$. Essentially, through \eqref{eqn:fdot_chirp}, the maximum $\fdot$, for a given chirp masses, fixes the maximum frequency that we can use to constrain $\ftilde$.

\begin{figure*}[ht!]
     \begin{center}
        \subfigure[ ]{%
            \label{fig:varyfdotmax}

        \includegraphics[width=0.5\textwidth]{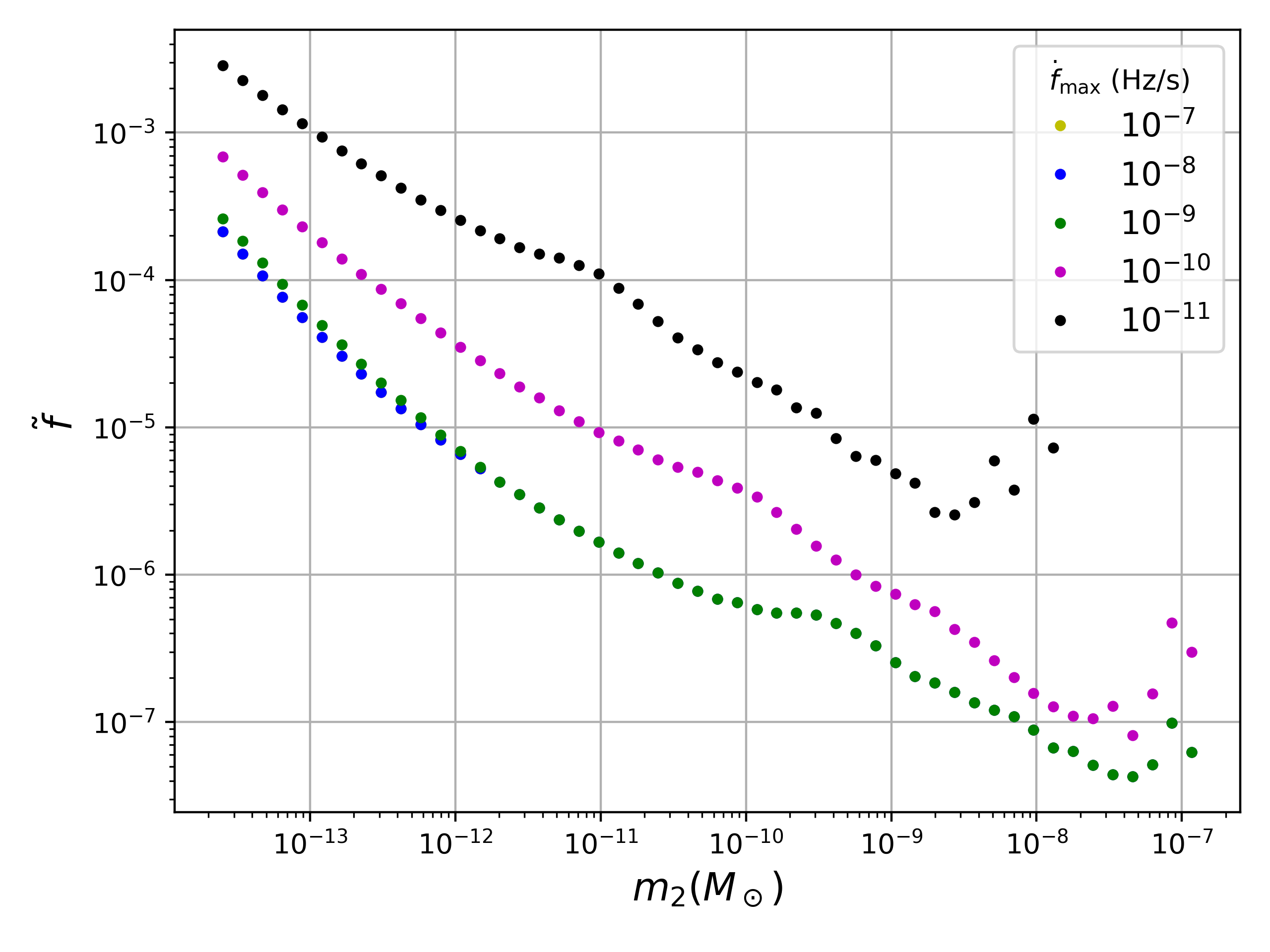}
        }%
        \subfigure[]{%
           \label{fig:1e-9vsnofdotmax}
           \includegraphics[width=0.5\textwidth]{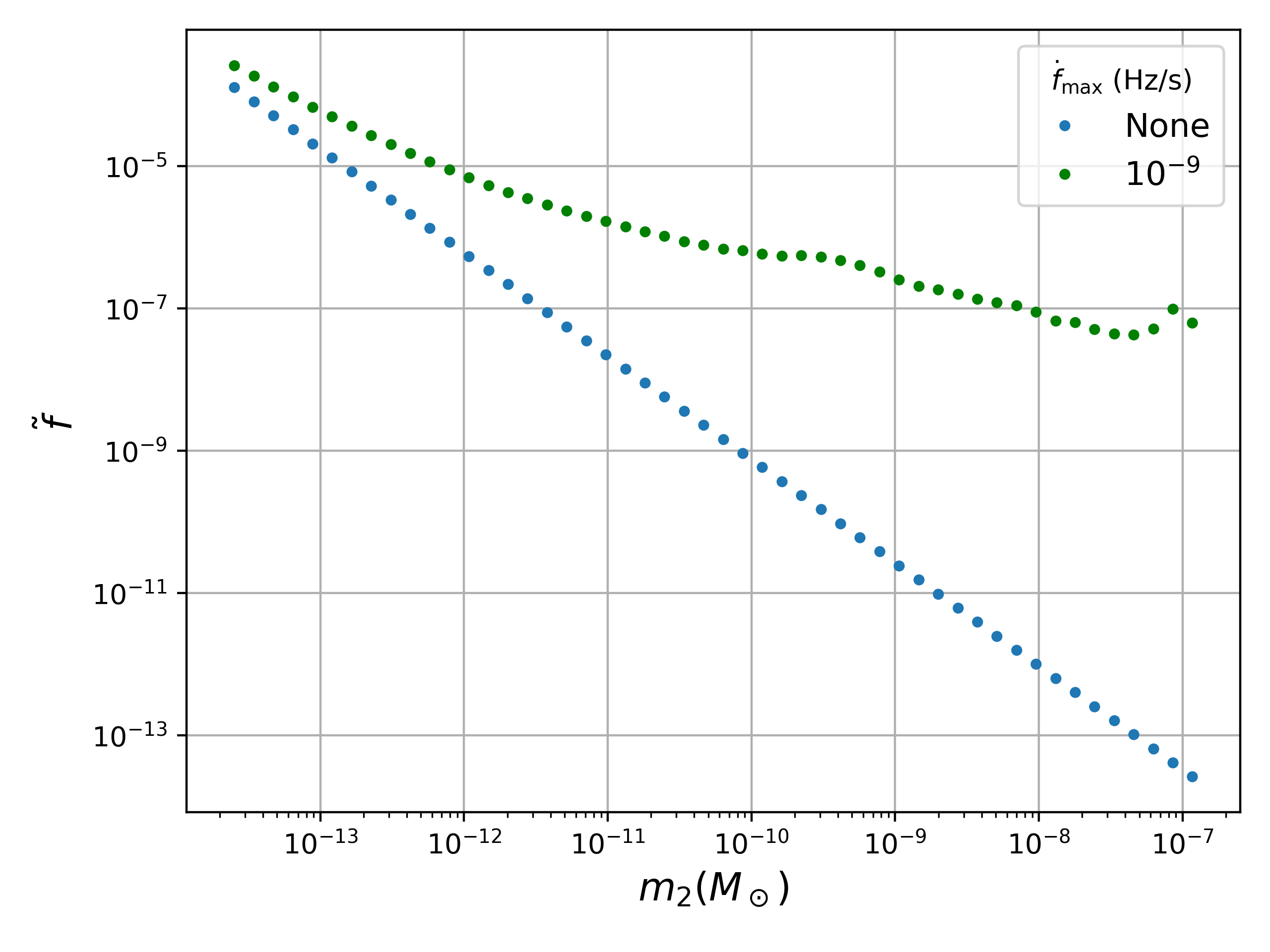}
        }\\ %  ------- End of the first row ----------------------%
    \end{center}
    \caption[]{Left: Varying the maximum spin-up to which \cwh searches are sensitive results in different constraints on $\ftilde$. Current searches consider $\fdotmax=10^{-9}$ Hz/s. We can see that smaller $\fdot$ indicates not only a poorer sensitivity at small $m_2$, but also an inability to reach higher values of $m_2$, since the signal spin up will increase to be higher than $\fdotmax$ during $\Tobs$. The degradation in sensitivity of smaller $\fdotmax$ occurs because signals at higher frequencies cannot contribute to the sum in \cref{eqn:ratedenssolved}, since the signal would either take on $\fdot\geq\fdotmax$, and/or the \gwh frequency evolution cannot be described by \cref{eqn:ftay} anymore. Right: A comparison showing how much the constraints would improve if no $\fdotmax$ existed, that is, if the signal could be searched for at arbitrarily high $\fdot$ with a frequency evolution following \cref{eqn:powlaws}. In both plots, we have set $\delta f =1$ Hz, which represents the approximate spacing in upper limits that is obtained through injections in \cwh searches \cite{KAGRA:2022dwb}. }%
     \label{fig:allvaryfdotmax}
\end{figure*}

We thus ask the question: if \cwh searches retain the signal model in \cref{eqn:ftay} but could increase $\fdotmax$, how would the ability to constrain $\ftilde$ change in \et? We provide an answer to this question in \cref{fig:allvaryfdotmax}. These plots show that for $10^{-9}$ Hz/s $\leq\fdotmax\leq 10^{-7}$ Hz/s, the projected constraints on $\ftilde$ are almost identical, since the linearity condition in \cref{eqn:ftay} is violated. Thus, if we remove both the linearity condition and $\fdotmax$, which would, in practice, require asteroid-mass \pbhs to be searched for with different methods \cite{Miller:2020kmv,Andrés-Carcasona:2023df,Alestas:2024ubs}, we find orders of magnitude improvement in the constraint on $\ftilde$. We will evaluate the suitability of these different methods to probe asteroid-mass \pbhs in \cref{subsec:change}.

\subsection{Impact of eccentricity}

Since we are considering binary black holes with such extreme mass ratios, we should evaluate to what extent our results are valid if these systems are eccentric. To begin, we evaluate the impact of eccentricity on the PN0 $\fdot$ term (\cref{eqn:fdot_chirp}) \cite{maggiore2008gravitational}:

\begin{align}
\fdot_{\rm ecc} &= \fdot_{\rm gw} g(\ecc) \\
g(\ecc) &= \left(1-\ecc^2\right)^{-7/2} \left(1 + \frac{73}{24}\ecc^2 + \frac{37}{96}\ecc^4\right)
\end{align}
where $g(\ecc)$ is a function arising from considering \gwh emission in a quasi-elliptical orbit of two masses, and $\ecc$ is the eccentricity of the system.
\tcr{The above equations describe how \gws are emitted more quickly with respect to those that arise from circular orbits, and come from an average over each orbit \cite{Peters:1963ux,Peters:1964zz}. They cannot account for any instantaneous variations in the time-frequency evolution \cite{Amaro-Seoane:2007osp}, such as harmonics \cite{Teukolsky:1973ha, Drasco:2005kz, Fujita:2009bp,Katz:2021yft}, bursts \cite{Glampedakis:2002cb, Rubbo:2006vh, Hopman:2006fc, Toonen:2009qi, Berry:2013poa, Oliver:2023xan}, transient resonances \cite{Pound:2007th,Barack:2009ux, Flanagan:2010cd, Gair:2010iv, Poisson:2011nh,Flanagan:2012kg, Berry:2016bit}, etc. In other words, these instantaneous frequency variations must be confined to one frequency bin in each $\TFFT$, and there can be no large-scale shift in the time-frequency evolution that deviates from \cref{eqn:ftay}. We thus only provide projected constraints in the regime in which such instantaneous, non-monotonic variations would be contained within a frequency bin and/or do not occur. }

In contrast to \mf, \cwh methods do not require phase coherence across the signal duration, but only within each $\TFFT$. Thus, if the spin-up induced by the eccentricity does not shift the signal frequency by more than one frequency bin in each $\TFFT$, \cwh methods would be sensitive to eccentric systems up to some value of eccentricity. In other words, we need to ensure the following condition is met

\begin{equation}
\fdot_{\rm ecc}\TFFT \leq \frac{1}{\TFFT}
\label{eqn:ecc-fdot-limit}
\end{equation}
so that \cref{eqn:ftay} is valid across the whole $\Tobs$.

\begin{figure}
    \centering
    \includegraphics[width=0.49\textwidth]{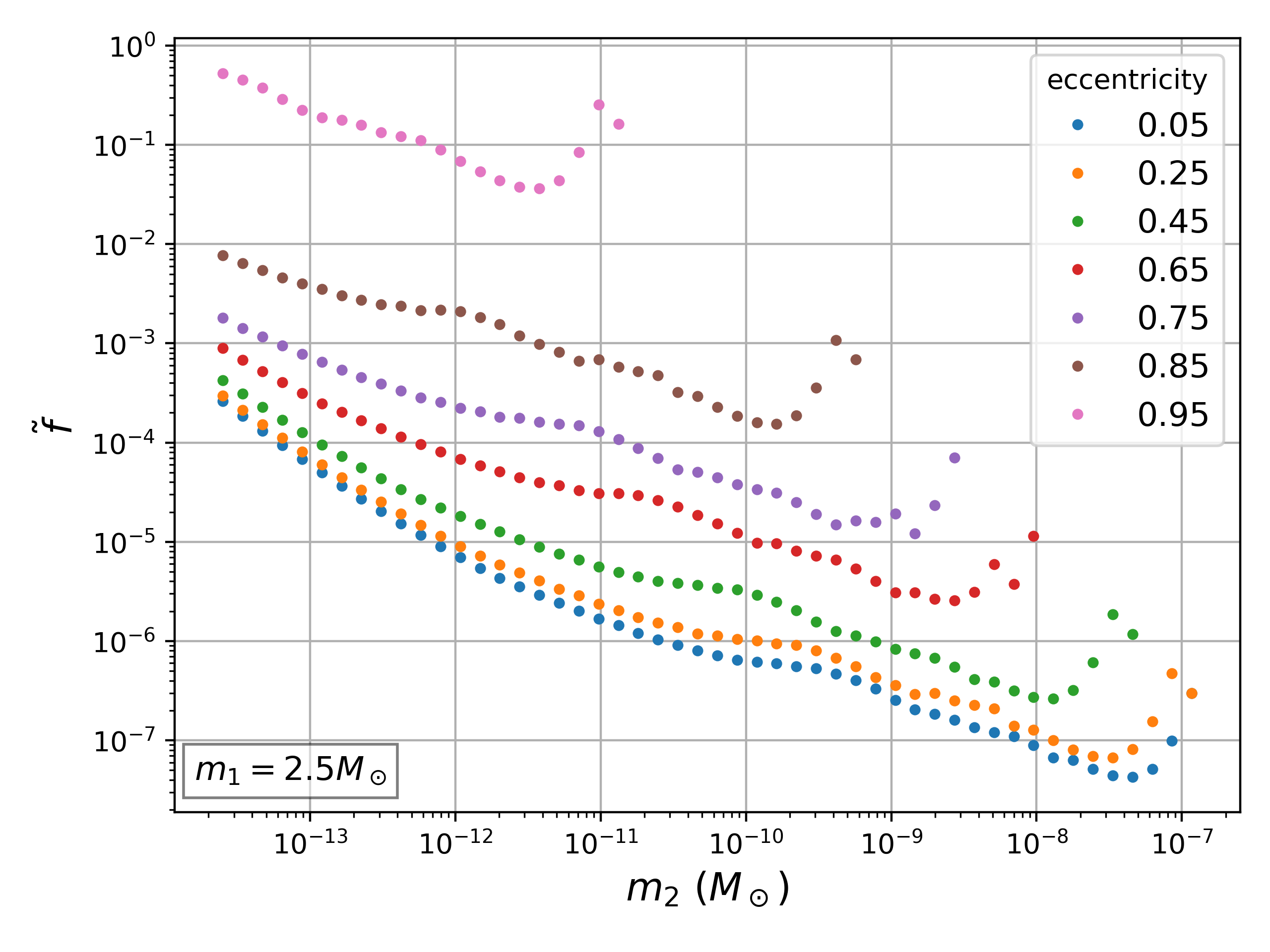}
    \caption{We vary the eccentricity and compute the expected constraint on $\ftilde$ for a fixed $m_1=2.5\msun$ \tcr{for a search that assumes $\fdotmax=10^{-9}$ Hz/s and using the \et \psd.} Highly eccentric systems are much harder to constrain with standard \cwh searches than lower ones, $\dot{f}_{\rm ecc}$ exceeds $\fdotmax$ much more easily than for low-eccentricity systems. }
    \label{fig:varyecc}
\end{figure}

How eccentric inspiraling systems are depends on how they formed \cite{Kowalska:2010qg,Samsing:2013kua,Samsing:2017xmd}. Additionally, only limited models exist for eccentricity evolution as a function of time \cite{Arredondo:2024nsl}. We thus assume the ``worst-case'' scenario in which the eccentricity is constant throughout the orbit, thus causing the maximum shift in $\fdot$. In \cref{fig:varyecc}, we show the impact of eccentricity on the projected constraints for $\ftilde$ for $m_1=2.5\msun$ using the \et sensitivity curve \tcr{and assuming $\fdotmax=10^{-9}$ Hz/s}. We follow the procedure outlined in \cref{sec:meth} to obtain the distance reached and thus the constraint on $\ftilde$.

In \cref{fig:varyecc}, we notice that as the eccentricity increases, the maximum $m_2$ probeable with \cwh methods decreases. This is because $\fdot_{\rm ecc}$ becomes too large and thus the condition in \cref{eqn:ecc-fdot-limit} is no longer satisfied. Furthermore, we see a general weakening of the constraint on $\ftilde$ as we increase eccentricity, since the systems with higher frequencies no longer contribute to the sum because the condition in \cref{eqn:ecc-fdot-limit} is more easily violated at high frequencies even though it is met at low frequencies.

\subsection{From which frequencies does the sensitivity come?}\label{subsec:sensfreq}

\begin{figure}
    \centering
    \includegraphics[width=0.49\textwidth]{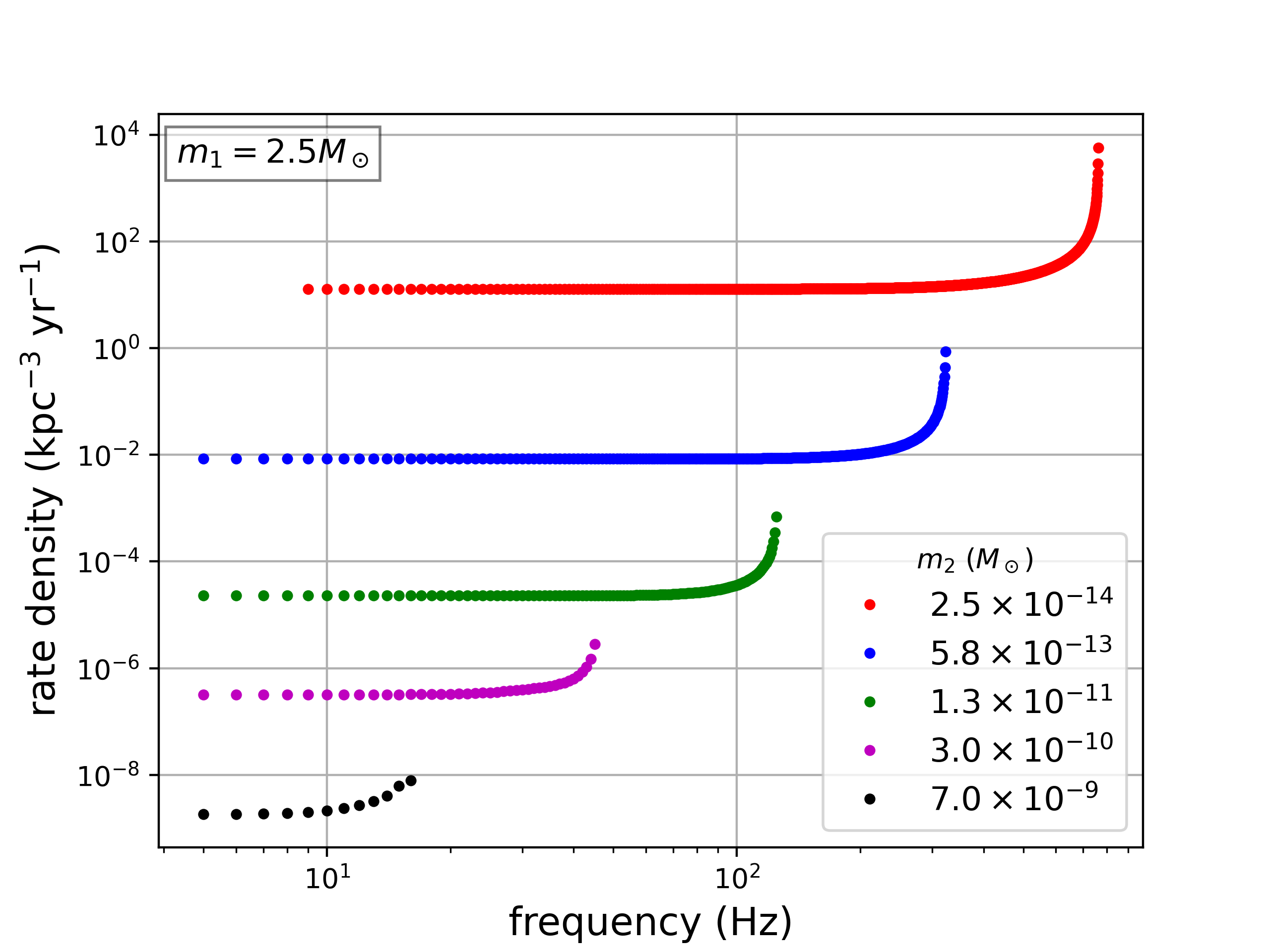}
    \caption{Accumulated rate density constraint as a function of which frequency we begin to sum the distance reaches in \cref{eqn:ratedenssolved} up to a maximum given by the right-most frequency on each curve. This has been computed using the \et \psd and assuming $m_1=2.5\msun$. The frequency on the x-axis represents the minimum frequency considered in the range $[\fmin,\fmax]$. For all values of $m_2$ plotted here, it is clear that the rate density saturates at a particular minimum frequency, which means that it is no longer beneficial to consider frequencies below $\fmin$ in the sum to compute constraints on the rate density.  }
    \label{fig:rate-dens-sum}
\end{figure}

Ironically, the restrictions on linearity in current CW searches imply that lower-frequency signals will always contribute to the sum, but they do not comprise a large gain in sensitivity with respect to those at high frequencies, thus motivating the need to attempt to detect signals at high frequencies and removing restrictions on $\fdotmax$ and linearity. We thus study specifically which distance reaches, indexed by the \gwh frequencies in the sum in \cref{eqn:ratedenssolved}, contribute the most to the rate density constraint. We show in \cref{fig:rate-dens-sum} constraints on the rate density for particular choices of $m_1$ and $m_2$, assuming negligible eccentricity, a linear frequency evolution (\cref{eqn:ftay}) and $\fdotmax=10^{-9}$ Hz/s. The way to interpret this plot is as follows: at a given frequency, the sum over distance reaches is taken from that frequency to the maximum one in the plot. For example, for the red curve, at $\fmin\simeq 200$ Hz, we sum the distance reaches from $\fmin\simeq 200$ Hz to $\fmax=800$ Hz. We notice immediately in the red curve that going to frequencies below 200 Hz does not improve the constraint on the rate density, indicating that the constraint on $\ftilde$ does not get better by looking at lower \gwh frequencies. We can see this behavior in each of the curves on this plot, except that the ``cut-off'' frequency below which it is no longer is useful to sum the distance reaches in \cref{eqn:ratedenssolved} decreases. This occurs because as $m_2$ increases, for a fixed $m_1$, $\Mc$ increases and thus $\fdot$ increases, indicating that, for higher frequencies, the \gwh frequency evolution will deviate from \cref{eqn:ftay}. We can see that for the highest value of $m_2$, that summing the distance reaches at all available frequencies is immensely helpful. 
% Additionally, \cref{fig:fmin-fmax-99perc} shows the frequency range $[\fmin,\fmax]$ necessary to analyze to obtain 99\% of the sensitivity on the rate density value, as a function of $m_2$. This plot is consistent with \cref{fig:rate-dens-sum}, showing that for smaller values of $m_2$, higher frequencies provide the best sensitivity.

% \begin{figure*}[ht!]
%      \begin{center}
% %
%         \subfigure[ ]{%
%             \label{fig:rate-dens-sum}

%         \includegraphics[width=0.5\textwidth]{rate_dens_vs_individ_summing_freqs_diff_m2.png}
%         }%
%         \subfigure[]{%
%            \label{fig:fmin-fmax-99perc}
%            \includegraphics[width=0.5\textwidth]{fmin_vs_fmax_to_get_99perc_rate_dens_lim.png}
%         }\\ %  ------- End of the first row ----------------------%
% %
%     \end{center}
%     \caption[]{\et sensitivity curve. $m_1=2.5\msun$. Left: Accumulated rate density constraint as a function at which frequency we begin to sum the distance reaches in \cref{eqn:ratedenssolved} up to a maximum given by the right-most frequency on each curve. Right: the frequency band to analyze to obtain 99\% of the optimal rate density constraint as a function of $m_2$, where the ``optimal'' constraint would arise if we sum distance reaches over the full frequency range.}%
%      \label{fig:sens-of-what-freqs}
% \end{figure*}

\subsection{Changes to current \cwh searches}\label{subsec:change}

Based on the discussion in \cref{subsec:sensfreq}, higher frequencies provide most of the constraining power on the rate density and $\ftilde$. In \cref{subsec:varyfdot}, we found that increasing $\fdotmax$ relative to that which is used currently ($\fdotmax=10^{-9}$ Hz/s) and searching for signals that evolve linearly does not alter the projected constraint on $\ftilde$ (\cref{fig:varyfdotmax}), but relaxing the linearity condition could provide orders of magnitude tightening of the constrains on $\ftilde$ (\cref{fig:1e-9vsnofdotmax}). We thus ask to what extent \emph{current} \cwh searches could constrain asteroid-mass \pbhs if these conditions on $\fdotmax$ and linearity are loosened, and what kinds of methods would be necessary to actually search for these systems.

\begin{figure}
    \centering
    \includegraphics[width=0.49\textwidth]{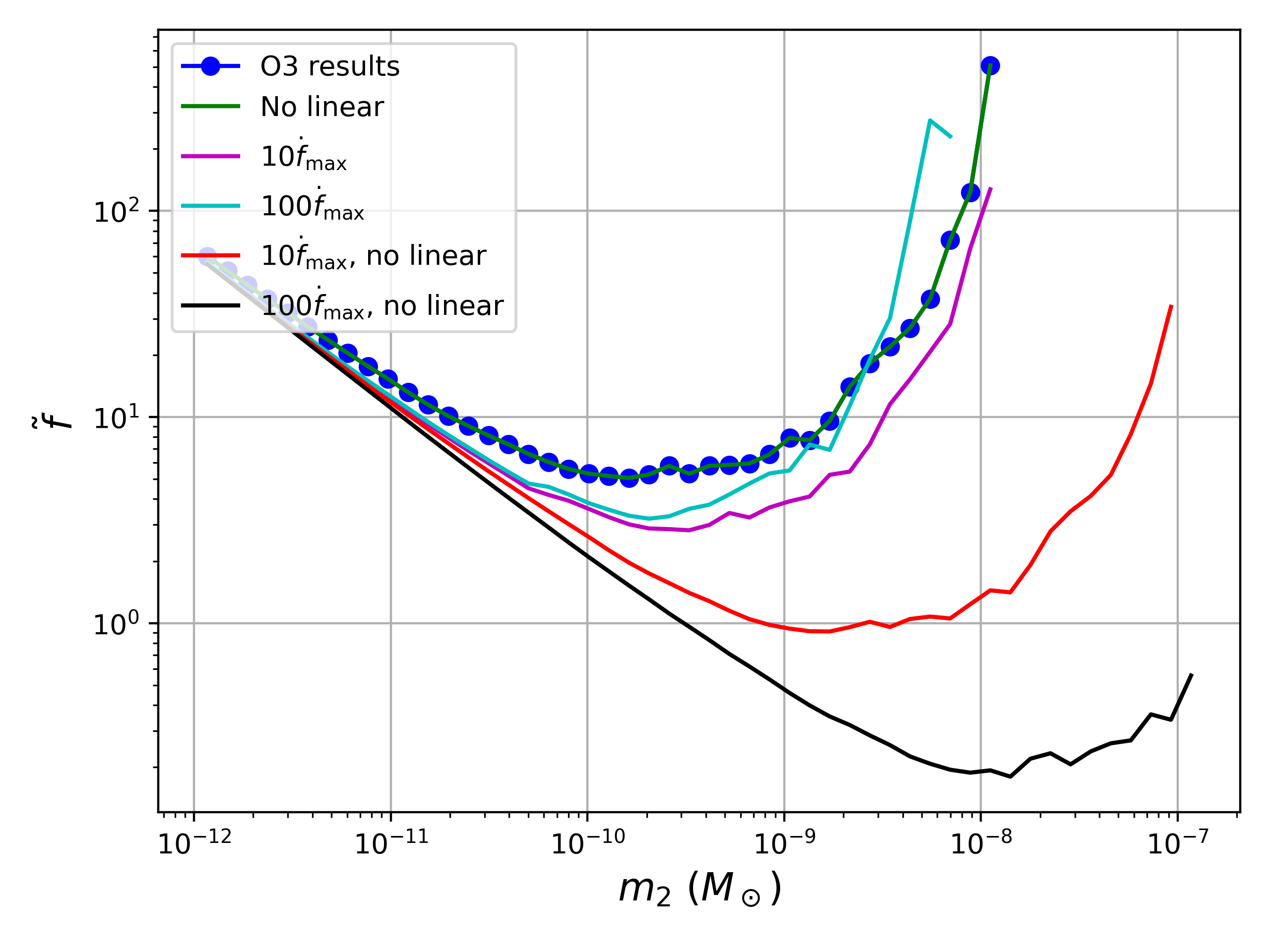}
    \caption{How the constraint on $\ftilde$ obtained in \cite{KAGRA:2022dwb} would change if various conditions of \cwh searches were relaxed. In particular, we compare results from O3 to the following conditions: (1) remove the requirement that the time-frequency evolution be linear; (2) increase $\fdotmax$ by a factor of 10 and 100, but still require the time-frequency evolution to be linear; and (3) again increase $\fdotmax$ by a factor of 10 and 100, but relax the requirement the time-frequency evolution to be linear. We can see that, even with data from the current observing run, if the \cwh methods could handle nonlinear time-frequency evolutions and also search up to a higher $100\fdotmax$, we would be able to constrain $\ftilde<1$ for $m_2\in [3\times 10^{-10},10^{-7}]\msun$. }
    \label{fig:o3-ideal}
\end{figure}

As an example of what could be achieved in the previous observing run of advanced \lvk (O3), we show in \cref{fig:o3-ideal} how the constraints on $\ftilde$ in \cite{KAGRA:2022dwb} would change if $\fdotmax=10^{-9}$ Hz/s were increased and if the linearity condition in \cref{eqn:ftay} were relaxed. In \cite{KAGRA:2022dwb}, $\delta f=\fdotmax\Tobs$ is used\footnote{This choice implies that, at each frequency at which an upper limit is set, that the signal frequency cannot vary by more than $\delta f$ over the course of $\Tobs$.}, so the spacing in the O3 interpolated upper limits changes when computing $\ftilde$ changes as a function of $\fdotmax$. Allowing $\fdotmax\rightarrow 100\fdotmax$ results in weaker constraints on $\ftilde$ than both the magenta and blue curves at certain masses. This occurs because, as shown in \cref{app:deltaf}, there is a non-trivial dependence of $\ftilde$ on the spacing in frequency $\delta f$ in the case of a flat \psd, in which larger $\delta f$ implies worse limits. However, the actual dependence of $\ftilde$ on $\delta f$, i.e. when including a frequency-dependent \psd and conditions on linearity and $\fdotmax$, result in a complicated behavior of these constraints, shown in \cref{fig:o3-ideal}.

\cref{fig:o3-ideal} implies that, even now, constraints of $\ftilde\lesssim 1$ could be obtained at masses below $\sim10^{-10}\msun$; however, relaxing the requirement in \cref{eqn:ftay} and allowing $100\fdotmax=10^{-7}$ Hz/s has implications for the computational cost of the search.
Increasing $\fdotmax$ would imply an increase in the size of the grid in $\fdot$ that is analyzed in \cwh searches. 
\tcr{Currently, the step size is \(\delta \dot{f} = 1/(\TFFT \Tobs) \sim 3 \times 10^{-11}\) Hz/s at high frequencies \cite{KAGRA:2022dwb}, where \(\TFFT = 1024\) s. To extend the positive \(\fdot\) coverage to \(\fdotmax = 10^{-8}\) Hz/s and \(\fdotmax = 10^{-7}\) Hz/s would require an additional 3,300 and 33,000 steps to be searched, respectively. This represents a 10x to 100x increase in the number of steps in spin-down, which currently includes approximately 320 negative steps and 32 positive steps at each sky position, which results in 10-100x more points in the parameter space to analyze.}
Such an increase in computational cost is not currently feasible, since all-sky \cwh searches already take $10^7$ CPU core-hours \cite{KAGRA:2022dwb,Miller:2024jpo}, and would not lead to a powerful constraint on $\ftilde$ at current detector sensitivity \emph{unless} the linearity condition was relaxed, which would, again, imply increased computational cost to consider $\ddot{f}$ or $\dddot{f}$ terms in \cref{eqn:ftay}. \tcr{However, searches that target a specific point in the sky, e.g. the \gc, could expand the range and number of spin-down parameters that are analyzed in order in order to probe less linear regimes for these signals.}

\section{Sensitivity of NEMO to asteroid-mass PBHs}\label{subsec:hfgw}

High-frequency gravitational-wave (\hfgw) detectors can deliver exquisite sensitivity in frequency ranges currently lacking by current and proposed ground-based detectors \cite{Aggarwal:2020olq}. In terms of inspiraling \pbhs, at such frequencies, the system will be extremely close to merger, and thus the purely \cwh approach discussed in \cref{sec:cwsearch} would break down. However, searches for ``transient \cws'' have been shown to be able to track rapid frequency variations of such inspiraling systems over time \cite{Miller:2020kmv,Andrés-Carcasona:2023df,Alestas:2024ubs,Miller:2024rca,Miller:2024fpo,Miller:2024jpo}, and even in the case of ``mini-\emri'' systems \cite{Guo:2022sdd}. While such searches, and also matched filtering ones, have not yet evolved to handle completely eccentric waveforms, it is worth considering how well such \emri systems could be constrained in future \hfgw detectors, in order to motivate the further development of these techniques.

\nemo \cite{Ackley:2020atn} is a planned \hfgw detector that plans to deliver exquisite sensitivity in the 2-4 kHz regime to increase the detection prospects of neutron star mergers, and is comparable to the sensitivities of \et and \ce. For our purposes, however, such high-frequency sensitivity simply implies that we could detect \pbh inspirals that are closer to merger than those we have considered in \cref{subsec:etsens}.

\begin{figure*}[ht!]
     \begin{center}
        \subfigure[ ]{%
            \label{fig:nemo-tcw}
        \includegraphics[width=0.5\textwidth]{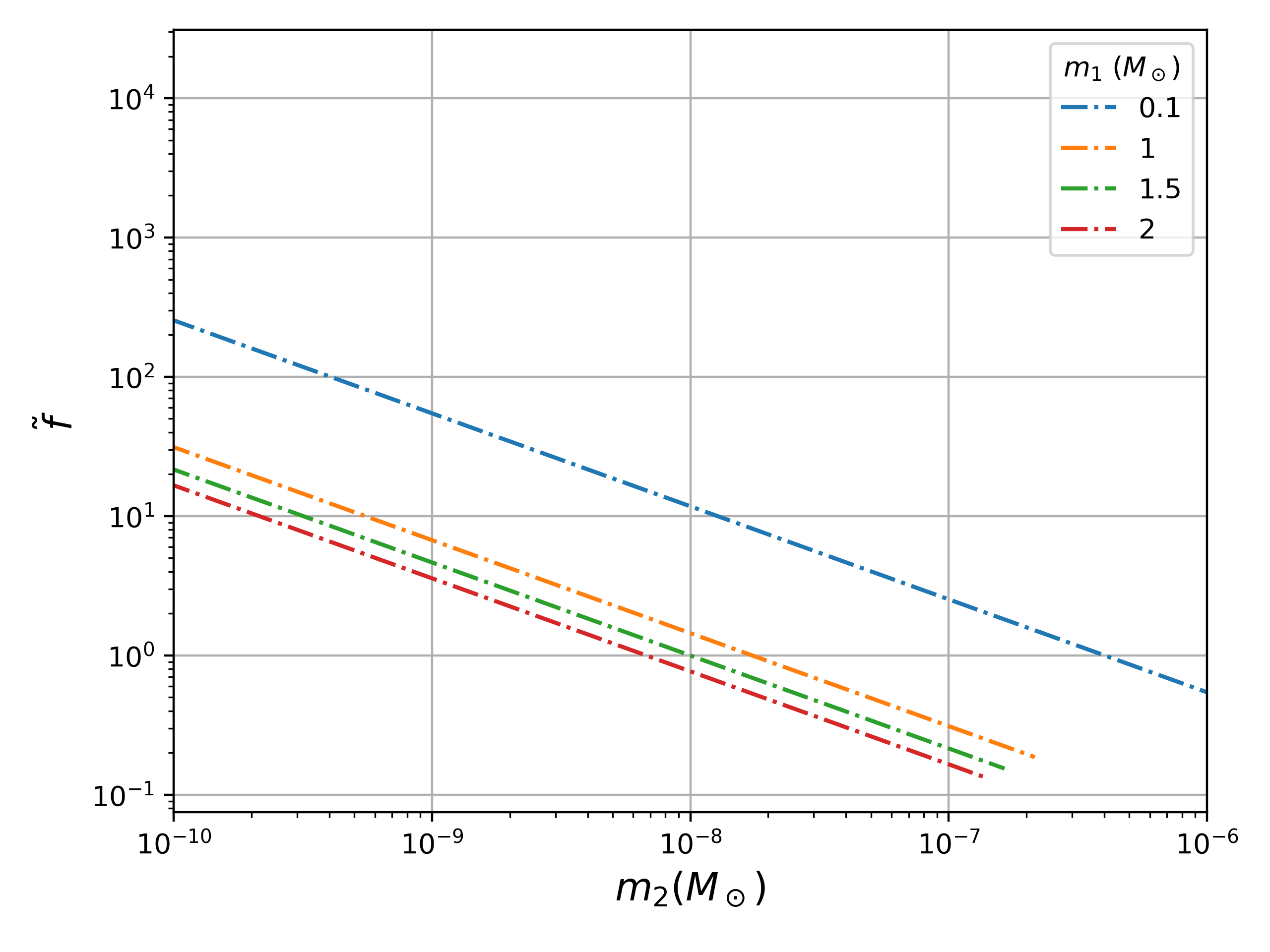}
        }%
        \subfigure[]{%
           \label{fig:nemo-mf}
           \includegraphics[width=0.5\textwidth]{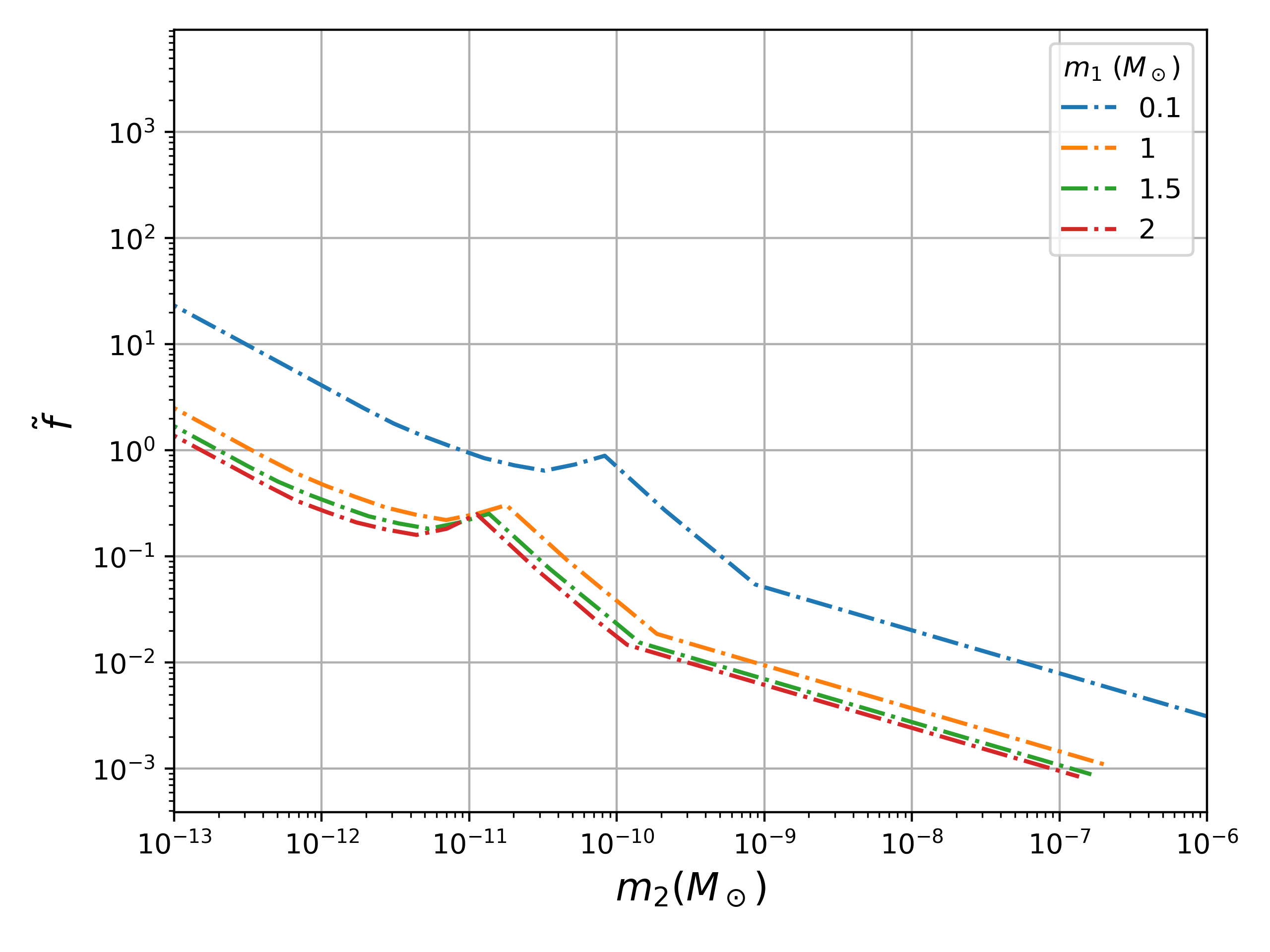}
        }\\ %  ------- End of the first row ----------------------%
    \end{center}
    \caption[]{Projected constraints on $\ftilde$ using a NEMO \psd between 1000 and 2500 Hz with two methodologies: (a) semi-coherent \cwh and (b) fully coherent \mf for different $m_1$. Coherent searches provide significantly stronger constraints at the expense of increased computational cost: \cwh methods can only constrain $\ftilde<1$ for $m_2>8\times 10^{-9}\msun$, while coherent \mf can constrain $\ftilde<1$ for $m_2>2\times 10^{-13}\msun$. In (b),two ``kinks'' appear: the first ($\sim 2\times 10^{-11}\msun$ for three of the curves, at $\sim 10^{-10}$ for the blue curve) when the signal duration becomes shorter than $\Tobs$, and the second ($\sim 10^{-10}\msun$ for three of the curves, $\sim 10^{-9}$ for the blue curve) when the frequency maximizing $d(f)$ shifts below 1000 Hz -- see text for details. }
     \label{fig:nemo}
\end{figure*}

Assuming a rough \psd value of $S_n(f)=10^{-48}$ strain/Hz and a frequency range of [1000,2500] Hz, we calculate the constraint on $\ftilde$ that would arise from a future analysis of NEMO data. We compute the projected constraints following the formalism presented in \cref{sec:meth}; however, because the \gwh signals become more transient-like and less \cwh-like as $m_2$ increases, we actually compute the distance reach using Eq. 32 in \cite{Miller:2020kmv}. Unfortunately, we only achieve $\ftilde\lesssim 1$ for $m_2\gtrsim10^{-8}\msun$ -- see \cref{fig:nemo-tcw}. Thus, we also compute the projected sensitivity assuming that we can use fully-coherent matched filtering. In this case, we calculate the luminosity distance reached following \cite{maggiore2008gravitational}:

\begin{equation}
    d = \frac{2}{5}\sqrt{\frac{5}{6}}\frac{c}{\pi^{2/3}}\left(\frac{G\Mc}{c^3}\right)^{5/6} \left(\int_{\fmin}^{\fmax} df \frac{f^{-7/3}}{S_n(f)}\right)^{1/2}\rho
\end{equation}
which, when evaluated in the case of $S_n(f)\sim$ constant:

% \begin{align}
% d &\simeq 3.87 \text{ kpc} \left(\frac{8}{\rho }\right)\left(\frac{\mathcal{M}} {10^{-5}M_\odot}\right)^{5/6} \nn \\ 
% &\times \left(\frac{1000\text{ Hz}}{\fmin}\right)^{2/3}\left(1-\left(\frac{\fmin/\fmax}{1.0001}\right)^{4/3}\right)
% \end{align}

\begin{align}
d &\simeq 3.87 \text{ kpc} \left(\frac{8}{\rho }\right)\left(\frac{\mathcal{M}} {10^{-5}M_\odot}\right)^{5/6}\left(\frac{1000\text{ Hz}}{\fmin}\right)^{2/3} \nn \\ 
&\times \left(\frac{1-\left(\fmin/\fmax\right)^{4/3}}{7.053\times 10^{-1}}\right) \left(\frac{10^{-48} \text{ Hz}^{-1} }{S_n(f)}\right)\label{eqn:dist-mf}
\end{align}
\tcr{This represents an idealized scenario with a detection threshold on the \snr $\rho=8$, consistent with many \mf searches.}
In \cref{fig:nemo-mf}, we show this projected constraint for different choices of $m_1$. 
% In the $m_1=0.1\msun$ curve, we observe a ``kink'' around $m_2\simeq 10^{-10}\msun$, which indicates a sort-of transition between when the signal duration exceeds the observation time, and when the signal becomes more ``transient-like''. 
We can observe two kinks in each curve in this plot. The first kink occurs when the signal duration no longer exceeds the observation time, i.e. it becomes more ``transient-like''\footnote{\tcr{Here, ``transient-like'' means that a signal lasts for a duration less than $\Tobs$: in particular, it could last as short as $\mathcal{O}($hours).}}. In practical terms, summing the contributions to the rate densities from different frequencies, outlined in \cref{sec:meth}, does not produce as competitive constraints as simply calculating the rate density in the Euclidean way (\cref{eqn:eucl}). The second kink separates two regimes of different slopes, and results because, for sufficiently light $m_2$, $\dmax$ occurs at a frequency between $[\fmin,\fmax]$. This happens because signals with these frequencies, in the span of $\Tobs$, would actually spin out of the $[\fmin,\fmax]$ band analyzed. Thus, we limit the frequency to which we allow to signal to spin up to $\fmax$, which corresponds to them lasting shorter than $\Tobs$. As $m_2$ increases, the frequency that maximizes $d(f)$ shifts to lower and lower frequencies. When that frequency falls below 1000 Hz, we observe the second kink, and the slope changes, since after the second kink, the maximum distance reach is always obtained by analyzing the full frequency range.

In practical terms, summing the contributions to the rate densities from different frequencies, outlined in \cref{sec:meth}, does not produce as competitive constraints as simply calculating the rate density in the Euclidean way:

\begin{align}
    \mathcal{R}&=\frac{3.0}{\avgVT} \\
    \avgVT &= \frac{4}{3}\pi d_{\rm max}^3\Tobs, \label{eqn:eucl}
\end{align}
where $\dmax$ and represents the system with a given $\Mc$ sweeping from $\fmin$ to $\fmax$ with a duration given by \cref{eqn:dt-bt-2-freqs}: $\Delta t<\Tobs$, i.e. a ``transient-like'' signal, not a \cwh. \tcr{For a chosen chirp mass, we calculate \cref{eqn:dist-mf} over many possible $[\fmin,\fmax]$ windows, and select the maximum of those to use in \cref{eqn:eucl}. }

\tcr{We note that the major difference between \cref{eqn:eucl} and \cref{eqn:ratedenssolved} is that the former applies in the case of signals whose durations are less then $\Tobs$, while the latter is applied for signals whose durations are greater than $\Tobs$. As explained in \cref{sec:meth}, \cref{eqn:ratedenssolved} allows us to account for the fact that multiple sources could be emitting \gws in the same frequency bin and thus enhance the signal strength. }

% We also considered the sensitivity of \lsd to detect \gws from asteroid-mass \pbhs. However, we have found that, even at design sensitivity and with matched filtering, we cannot provide $\ftilde<1$ unless the noise amplitude spectral density is improved by about an order of magnitude.

% \subsection{LSD}

% \lsd is being designed to detect \gws at $\sim 10$ kHz.

% We have found that matched-filtering cannot make any meaningful statement on $\ftilde$ to detect inspiraling \pbhs using \lsd with the expected noise amplitude spectral density $\sqrt{S_n(f)}=10^{-23}$. Thus, we have varied $\sqrt{S_n(f)}$ to understand what level of the noise would be necessary to probe $\ftilde\lesssim 1$ between $[10^{-13},10^{-11}]\msun$, and have found that $\sqrt{S_n(f)}\lesssim 10^{-24.5}\unitasd$ is necessary to begin to obtain a physical constraint in this regime.

% \begin{figure}[ht!]
%      \begin{center}
%            \label{fig:vary-sn-lsd}
%            \includegraphics[width=0.49\textwidth]{m2_vs_ftilde_vary_sn_lsd_cbc.png}
%     \end{center}
%     \caption[]{Projected constraints on $\ftilde$ using the \lsd experiment with matched filtering. We fix $m_1=0.1\msun$ and vary $S_n(f)$. We again observe two kinks, whose explanations are the same as those in \cref{fig:nemo-mf}.   }%
%      \label{fig:all-lsd}
% \end{figure}

\section{Conclusions}

In this paper, we have shown that \cwh and transient \cwh searches could constrain the existence of asteroid-mass \pbhs in \emri systems currently and with future \gwh detectors. In the mass regime in which our projected constraints overlap with those from microlensing, our results provide complementary ways of probing \pbhs that could have formed in binary systems, instead of isolated ones. Additionally, in the so-called ``asteroid-mass'' regime, our results indicate that \gwh detectors would provide the first-ever stringent constraints on the fraction of \dm that \pbhs could compose. We note that we have parameterized our constraints in terms of $\ftilde$ in order to remain model-agnostic, and therefore in order to directly and fairly compare with the $\fpbh$ limits that arise from microlensing experiments or theoretical considerations, we would have to know exactly which assumptions are made on the \pbh mass functions and formation mechanisms.

In addition to providing projected constraints, we have evaluated how these constraints would change if we allow the binaries to take on nonzero eccentricities, if we change the maximum $\fdot$ up to which is searched, and if we relax the requirement that the \gwh signal be quasi-monochromatic. Our results show that eccentricity plays a major role in affecting our constraints, and that incorporating it into an analysis may be necessary to achieve the best possible constraints. Furthermore, we determine which \gwh frequencies we should analyze as a function of the \pbh mass $m_2$ in order to determine where most of the constraining power lies. We find that higher- frequencies for very small $m_2$ values contains most of the constraining power, while lower frequencies are necessary for heavier $m_2$, since the requirement on $\fdotmax$ prohibits higher frequencies from contributing to the sum in \cref{eqn:ratedenssolved}. This study has implications for all-sky searches for \pbh inspirals: in fact, we could envision a search in which higher frequencies are prioritized for smaller systems, while lower frequencies are analyzed for heavier ones, instead of blindly searching for all systems at all frequencies. Such a scheme could reduce the computational burden of an all-sky search for \pbh inspirals.

Our results show that \cwh search techniques, exactly as they are, could provide stringent constraints on $\ftilde$ in the \et era of \gwh detectors. \tcr{However, future work will entail understanding to what extent \tcw methods can address the nonlinear signal frequency evolution in \cref{eqn:powlaws}. In this case, the signal may even last much shorter than, depending on the component masses and initial orbital separation, and thus requires an algorithm capable of efficiently tracking curved, non-stationary trajectories in the time-frequency plane. Multiple methods exist to perform this kind of time-frequency tracking \cite{Miller:2018rbg,Sun:2018hmm,Oliver:2018dpt,Banagiri:2019obu,Keitel:2019zhb,Miller:2020kmv,Miller:2024jpo,Alestas:2024ubs}; however, their computational cost, sensitivity and sky coverage must be evaluated in order to assess to what extent they could be sensitive to asteroid-mass \pbhs now and in next-generation \gwh \ifos.}

Our results thus demonstrate that \cwh search techniques, when adapted to capture the distinctive frequency evolution of asteroid-mass \pbh inspirals, can provide a powerful probe using both current and future \gwh observatories. This is in line with the adaption of \cwh methods to search for a variety of \dmh signatures \cite{DAntonio:2018sff,Pierce:2018xmy,Palomba:2019vxe,Grote:2019uvn,Guo:2019ker,Armaleo:2020efr,Vermeulen:2021epa,Miller:2022wxu,Manita:2023mnc,LIGOScientific:2021odm,KAGRA:2022dwb,Gottel:2024cfj} -- see \cite{Miller:2025yyx} for a recent review. While next-generation detectors such as \et and \nemo will open up vast new regions of parameter space, even modest modifications to present-day search strategies could already yield meaningful constraints on the fraction of \dm that \pbhs can compose. Taken together, these findings highlight the untapped potential of \cwh methods to test scenarios for \pbhs that are otherwise difficult to access, and they motivate the development of more flexible search pipelines capable of capturing the full diversity of long-lived, \tcw signals.

% Furthermore, if \tcw methods are used now, they could provide tight constraints on $\ftilde$ using O3 and subsequent observing runs of advanced \lvk, highlighting both the short-term and long-term impact of \cwh searches for asteroid-mass \pbhs.

\appendix

% % \section{Additional equations}\label{app:sens}

% We provide here more information about \cref{eqn:h0min}, in particular the parameter $\Lambda$ that encapsulates particular analysis choices made in a real search for quasi-monochromatic \gws. It can be written as \cite{Astone:2014esa}

% \begin{align}
%     \Lambda &= 4.02\left(\frac{p_0(1-p_0)}{\thetathr^2 p_1^2}\right)^{1/4} \times \sqrt{CR_\text{thr}-\sqrt{2}\erfc^{-1}(2\Gamma)} \nn \\ 
%     p_0&=e^{-\thetathr}-e^{-2\thetathr}+\frac{1}{3}e^{-3\thetathr} \nonumber \\
% p_1 &=  e^{-\thetathr} - 2e^{-2\thetathr} + e^{-3\thetathr}
% \end{align}
% where, typically, $\thetathr$ is the threshold to determine which time-frequency pixels in the equalized power spectrogram are important, $\Gamma$ is the chosen confidence level, and $CR_\text{thr}$ is a threshold on a detection statistic called the critical ratio, which corresponds to the number of standard deviations a particular outlier is away from the mean. When $\thetathr=2.5$, $\Gamma=0.95$ and $CR_\text{thr}$=5, $\Lambda=12.81$.

\section{Dependence of constraints on $\delta f$}\label{app:deltaf}

We mentioned that the constraint on $\ftilde$ has some dependence on the choice we make for $\delta f$ in the context of the results from O3 show in \cref{fig:o3-ideal}. Here, we evaluate how $\delta f$ could affect the sensitivity of the search. We note, however, that we consider a very simplified case in which we have a flat noise \psd, in order to arrive at a semi-analytical expression for the detectable space-time volume $\avgVT$, which then allows us to constrain the rate density and $\ftilde$.

We know that $\avgVT$ depends on $\dmax^3$ and $T$ through \cref{eqn:vt}, and we will work in the small $\delta f$ limit, that is $\delta f\ll f$.
We can write an abbreviated form of \cref{eqn:h0}:
\begin{align}
    d(f)&\propto f^{2/3} \nn \\  
    T &\propto \delta f f^{-11/3} \nn \\
    V &\propto d(f)^3 \propto f^2 \nn \\
     \avgVT &\propto \delta f f^{-5/3} 
\end{align}
We thus see how $\avgVT$ depends on $f$ and $\delta f$. Now, we note that when we use \cref{eqn:ratedenssolved}, we are actually summing the distances at particular frequencies in steps of $\delta f$, since, typically, the upper limits on $h_{0,\rm min}$ are given in steps of $\delta f$. In mathematical terms, this corresponds to a summation of the form:

\begin{align}
     \avgVTtot &= \sum \avgVT \propto \delta f \sum_{n=0}^{N} f_n^{-5/3} \nn \\ 
     &\propto \delta f (\fmin^{-5/3} +(\fmin+\delta f)^{-5/3}+ (\fmin+2\delta f)^{-5/3} \\ &+ ...) \nn \\ 
     &\propto \delta f\fmin^{-5/3}\sum_{n=0}^{N}\left(1+n\frac{\delta f}{\fmin}\right)^{-5/3} 
\end{align}
where $N=(\fmax-\fmin)/\delta f$ is the number of frequencies that we sum.
As noted before, for simplicity, we have set $S_n(f)=$ constant, but its variation with frequency will affect the sum in practice, and could be parameterized as ``weights'' that affect each term in the sum.

This particular sum can in fact be represented in terms of Hurwitz zeta functions:

\begin{equation}
    \sum_{n=0}^N (1+nz)^{-5/3} = 1+\frac{\zeta(\frac{5}{3}, 1 + \frac{1}{z}) - \zeta(\frac{5}{3}, N + \frac{1}{z} + 1)}{z^{5/3}}
    \label{eqn:onlysum}
\end{equation}
where $z=\delta f/\fmin$, and when applied to our case:

% \begin{widetext}
% \begin{equation}
%     \avgVT_{\rm tot}\propto \delta f^{-2/3}\left\{\left(\frac{\delta f}{\fmin}\right)^{5/3}+\left[\zeta\left(\frac{5}{3}, 1 + \frac{\fmin}{\delta f}\right) - \zeta\left(\frac{5}{3}, N + \frac{\fmin}{\delta f} + 1\right)\right]\right\}\label{eqn:zeta-spec}
% \end{equation}
% \end{widetext}
% and with the other variables:

\begin{widetext}
\begin{align}
    \avgVTtot &=  \frac{5}{96\pi^{2/3}} \frac{1}{\Lambda^3} \left(\frac{\TFFT^{1/2} \Tobs^{1/2}}{{S_n}}\right)^{3/2}\left(\frac{G \mathcal{M}}{c^{21/10}}\right)^{10/3} \delta f^{-2/3}\left\{\left(\frac{\delta f}{\fmin}\right)^{5/3}+\left[\zeta\left(\frac{5}{3}, 1 + \frac{\fmin}{\delta f}\right) - \zeta\left(\frac{5}{3}, N + \frac{\fmin}{\delta f} + 1\right)\right]\right\}\label{eqn:zeta-spec-complete}
\end{align}
\end{widetext}
where we have dropped the $f$ argument on $S_n$ to indicate that we have considered a flat noise spectral density.

\begin{figure}
    \centering
    \includegraphics[width=0.49\textwidth]{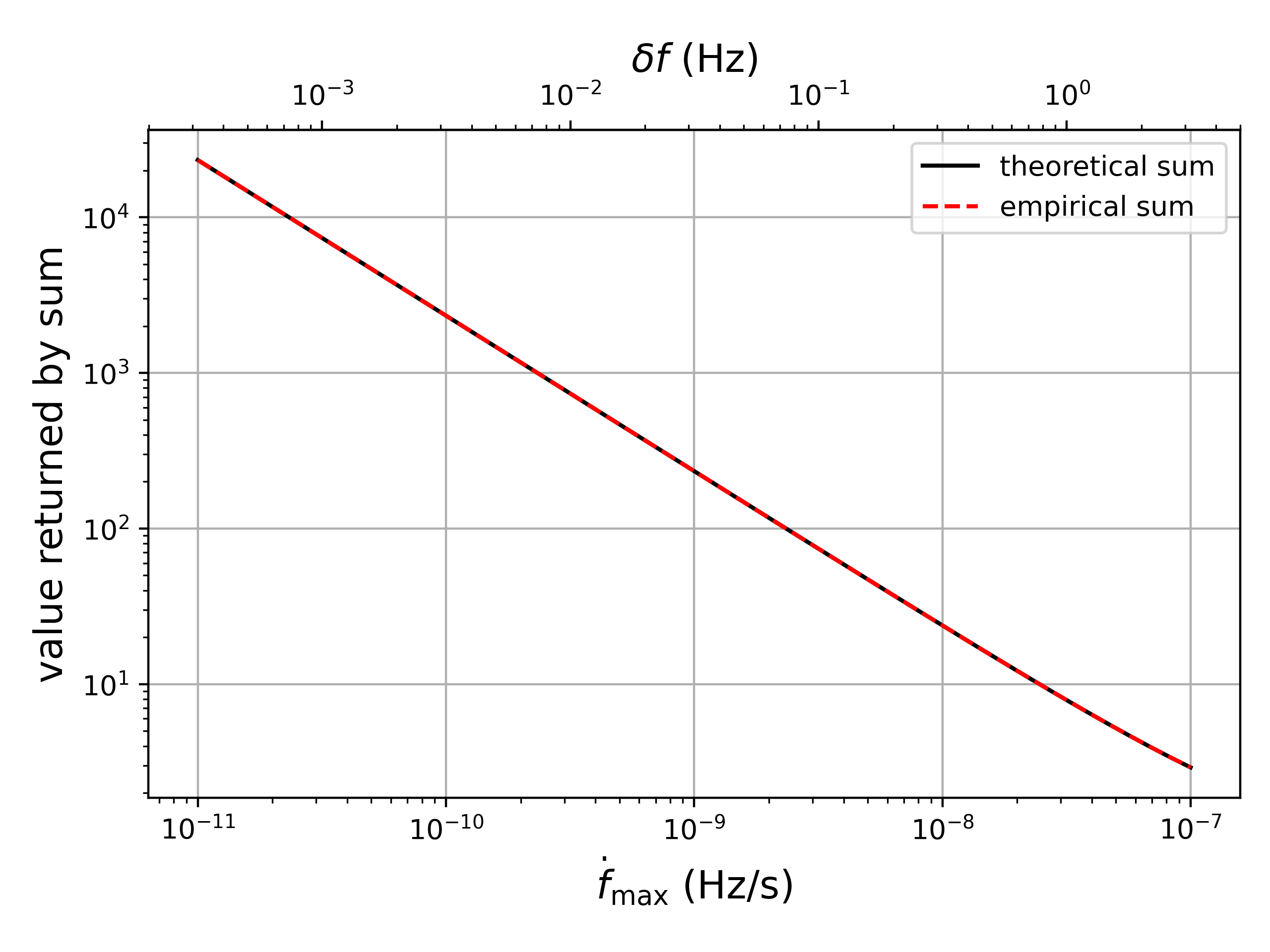}
    \caption{Comparison of \cref{eqn:onlysum} and \cref{eqn:ratedenssolved}, where in \cref{eqn:ratedenssolved}, we have neglected the prefactors and divided out the common frequency $\fmin^{-5/3}$.}
    \label{fig:sum-compare}
\end{figure}

We will fix $\fmin=5$ Hz and $\fmax=2000$ Hz, and consider to what extent this formulation agrees with what we actually do in \cref{eqn:ratedenssolved}, and how $\avgVT$ depends on $\delta f$. We show this comparison in \cref{fig:sum-compare}, and conclude that our theoretical formulation matches with the empirical one.

\begin{figure}
    \centering
    \includegraphics[width=0.49\textwidth]{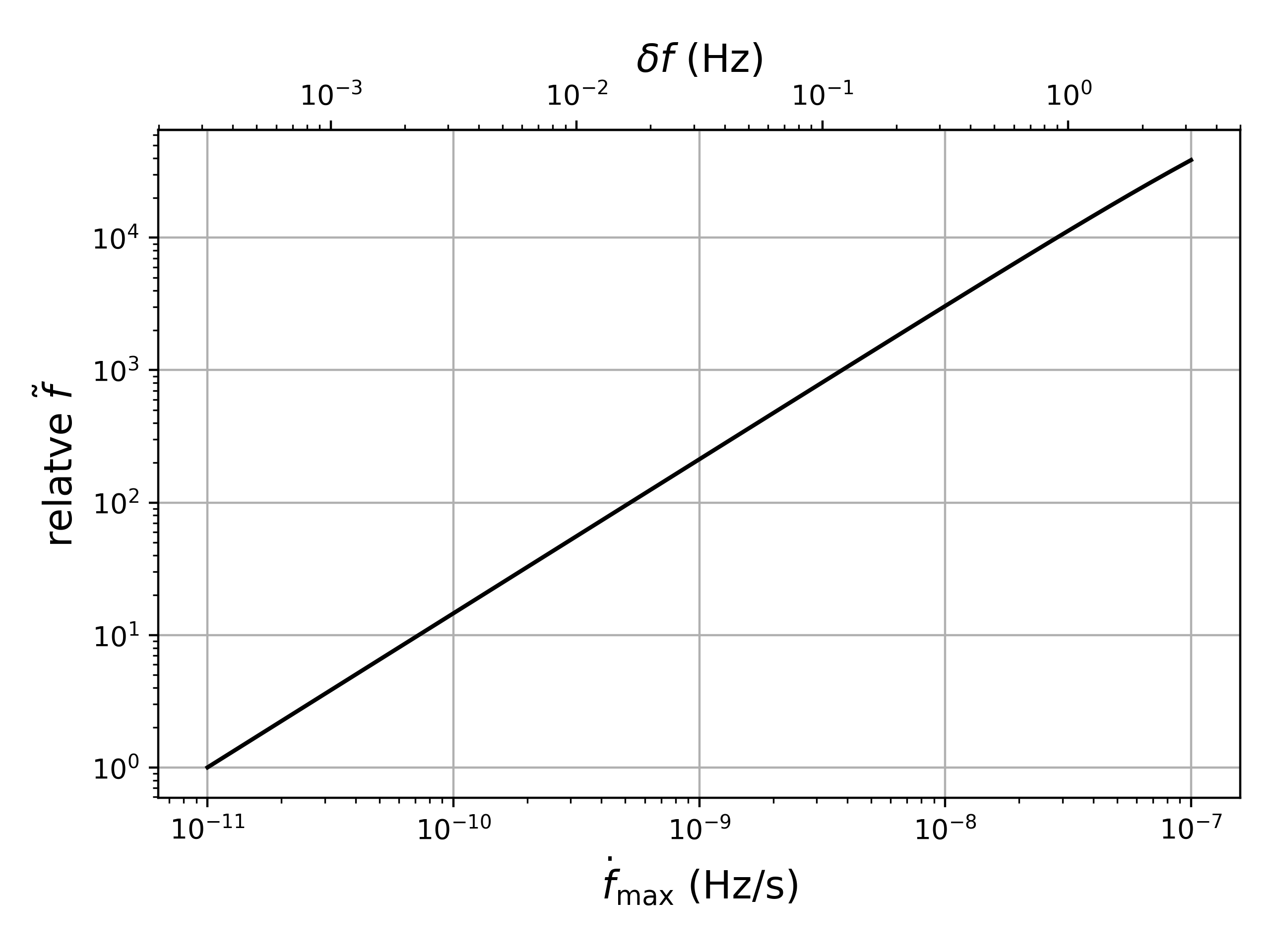}
    \caption{The increase in the constraint $\ftilde$ with respect to the minimum $\ftilde$ attainable in this example with the smallest $\delta f$ spacing. Note that this figure assumes a flat \psd and no restrictions on linearity or $\fdotmax$. In practice, the relative $\ftilde$ is much smaller.}
    \label{fig:ftilde-worse}
\end{figure}

We then compute the impact of $\delta f$ on $\ftilde$:

\begin{align}
    \mathcal{R}&\propto \avgVTtot^{-1} \\
    \tilde{f}^{53/37} &\propto \avgVTtot^{-1} \\
    \tilde{f} &\propto \avgVTtot^{-37/53}
\end{align}
and show the ratio of $\ftilde$ to $\ftilde_{\rm min}$ in \cref{fig:ftilde-worse}.

From \cref{eqn:zeta-spec-complete} and from \cref{fig:sum-compare}, we see a quite complicated dependence on $\delta f$, and note that what we have just derived above is \emph{theoretical}, assuming a fixed noise \psd. The real situation is more complicated, and we cannot find a closed-form expression for the sum, since, in the case of a varying \psd, the sum on the left-hand side in \cref{eqn:onlysum} takes on frequency-dependent weights. Additionally, the theoretical formulation does not impose any restrictions on linearity or on the maximum $\fdot$ that a \gwh signal could take, both of which affect the constraint on $\ftilde$. However, our theoretical results highlight that there \emph{is} a dependence on on the spacing in frequency $\delta f$ of the upper limits $h_0(f)$, which explains why, in \cref{fig:o3-ideal}, that changing $\fdotmax$ does not necessarily lead to better constraints on $\ftilde$.

\section*{Acknowledgments}
This material is based upon work supported by NSF's LIGO Laboratory which is a major facility fully funded by the National Science Foundation

This research has made use of data, software and/or web tools obtained from the Gravitational Wave Open Science Center (https://www.gw-openscience.org/ ), a service of LIGO Laboratory, the LIGO Scientific Collaboration and the Virgo Collaboration. LIGO Laboratory and Advanced LIGO are funded by the United States National Science Foundation (NSF) as well as the Science and Technology Facilities Council (STFC) of the United Kingdom, the Max-Planck-Society (MPS), and the State of Niedersachsen/Germany for support of the construction of Advanced LIGO and construction and operation of the GEO600 detector. Additional support for Advanced LIGO was provided by the Australian Research Council. Virgo is funded, through the European Gravitational Observatory (EGO), by the French Centre National de Recherche Scientifique (CNRS), the Italian Istituto Nazionale della Fisica Nucleare (INFN) and the Dutch Nikhef, with contributions by institutions from Belgium, Germany, Greece, Hungary, Ireland, Japan, Monaco, Poland, Portugal, Spain.

We would like to thank all of the essential workers who put their health at risk during the COVID-19 pandemic, without whom we would not have been able to complete this work.

\bibliographystyle{apsrev4-1}
\bibliography{biblio,biblio_pbh_method}

\end{document}